\documentclass[twocolumn,trackchanges]{aastex62}
\usepackage{natbib}
\usepackage{amsmath}
\usepackage{amssymb}
\hypersetup{linkcolor=red,citecolor=orange,filecolor=cyan,urlcolor=magenta}

\newcommand{\ang}{\mbox{\AA} }

\newcommand{\galex}{\textit{GALEX}}
\newcommand{\hst}{\textit{HST}}
\newcommand{\spitzer}{\textit{Spitzer}}
\newcommand{\herschel}{\textit{Herschel}}
\newcommand{\magphys}{\texttt{MAGPHYS}}
\newcommand{\magphysz}{\texttt{MAGPHYS+photo}-$z$}

\newcommand{\tauv}{\hbox{$\hat{\tau}_{V}$}}

\shortauthors{Battisti et al.}

\begin{document}

\title{The Strength of the 2175\ang Feature in the Attenuation Curves of Galaxies at $0.1< z\lesssim3$}

\correspondingauthor{A. J. Battisti}
\email{andrew.battisti@anu.edu.au}

\author[0000-0003-4569-2285]{A. J. Battisti}
\affil{Research School of Astronomy and Astrophysics, Australian National University, Cotter Road, Weston Creek, ACT 2611, Australia}
\affil{ARC Centre of Excellence for All Sky Astrophysics in 3 Dimensions (ASTRO 3D), Australia}

\author[0000-0001-9759-4797]{E. da Cunha}
\affil{Research School of Astronomy and Astrophysics, Australian National University, Cotter Road, Weston Creek, ACT 2611, Australia}
\affil{International Centre for Radio Astronomy Research, University of Western Australia, 35 Stirling Hwy, Crawley, WA 6009, Australia}
\affil{ARC Centre of Excellence for All Sky Astrophysics in 3 Dimensions (ASTRO 3D), Australia}

\author[0000-0003-4702-7561]{I. Shivaei}
\affil{Steward Observatory, University of Arizona, 933 N Cherry Ave, Tucson, AZ 85721, USA}
\affil{Hubble Fellow}

\author[0000-0002-5189-8004]{D. Calzetti}
\affil{Department of Astronomy, University of Massachusetts-Amherst, Amherst, MA 01003, USA}
\collaboration{(COSMOS collaboration)}

\begin{abstract}
We update the spectral modeling code \magphys\ to include a 2175\ang absorption feature in its UV-to-near-IR dust attenuation prescription. This allows us to determine the strength of this feature and the shape of the dust attenuation curve in $\sim$5000 star-forming galaxies at $0.1< z\lesssim3$ in the COSMOS field. We find that a  2175\ang absorption feature of $\sim$1/3 the strength of that in the Milky Way is required for models to minimize residuals. We characterize the total effective dust attenuation curves as a function of several galaxy properties and find that the UV slopes of the attenuation curve for COSMOS galaxies show a strong dependence with star formation rate (SFR) and total dust attenuation ($A_V$), such that galaxies with higher SFR and $A_V$ have shallower curves and vice versa. These results are consistent with expectations from radiative transfer that attenuation curves become shallower as the effective dust optical depth increases. We do not find significant trends in the strength of the 2175\ang absorption feature as a function of galaxy properties, but this may result from the high uncertainties associated with this measurement. The updated code is publicly available online. 
\end{abstract}

\section{Introduction}
Interstellar dust within galaxies plays a dominant role in altering their observed spectral energy distributions (SEDs) by acting to obscure light from ultraviolet (UV) to near-infrared (NIR) wavelengths and reemitting it at infrared (IR) wavelengths. The wavelength-dependent behavior of dust attenuation, referred to as the attenuation curve, in each galaxy depends strongly on the geometry of the dust relative to the stellar distribution \citep[see review by][]{calzetti01} and therefore it is expected to vary considerably on a galaxy-by-galaxy basis, even in situations where the intrinsic dust properties (characterized by the dust extinction curve) between galaxies is similar\footnote{We define attenuation to be a combination of extinction, scattering of light into the line of sight by dust, and geometrical effects due to the star-dust geometry. Extinction is the absorption and scattering of light out of the line of sight by dust, which has no dependence on geometry. Unlike extinction curves, the geometric effects in attenuation make it difficult to interpret the physical properties of dust from attenuation curves.} \citep[e.g.,][]{narayanan18}. Accurately characterizing the variation of dust attenuation curves is important because the assumed shape of dust curves have a strong impact on the physical properties derived from SED modeling \citep{conroy13} and on the accuracy of photometric distance (photo-$z$) estimates \citep{battisti19}. 

A common characteristic of dust extinction and attenuation curves is that the strongest extinction/attenuation typically occurs in the UV (bluer wavelengths) and decreases towards the NIR (redder wavelengths) in a manner that is gradual (featureless), with the possible exception of a broad absorption feature centered at 2175\ang \citep[referred to as the 2175\ang feature or bump; e.g.,][]{draine03}. This feature is observed in many sightlines of the Milky Way \citep[MW;][]{cardelli89, fitzpatrick99}, the Large Magellanic Cloud \citep[LMC; e.g.,][]{gordon03}, and the Andromeda galaxy \citep[M31; e.g.,][]{bianchi96,clayton15}, but is typically very weak or absent in the Small Magellanic Cloud \citep[SMC; e.g.,][]{gordon03}. Theoretical studies have shown that the apparent strength of the 2175\ang feature in attenuation curves, relative to the intrinsic extinction curve, can be considerably reduced due to the additional geometric and scattering effects at play \citep[e.g.,][]{gordon97, witt&gordon00, seon&draine16}. 

Numerous observational studies have looked into the strength of the 2175\ang feature in the attenuation curves of galaxies, often with noticeably different results. \citet{calzetti94} found that starburst galaxies (strongly star-forming galaxies) lack the 2175\ang feature entirely in their attenuation curve, whereas studies of local ``normal" star-forming galaxies (SFGs) suggest a weakened 2175\ang feature is present in a fraction of galaxies \citep[e.g.,][]{conroy10, wild11, battisti17b, salim18}. Similar amounts of variation are observed at higher redshifts, with some studies favoring the inclusion of a weak feature \citep[e.g.,][]{noll09a, buat11b, buat12, kriek&conroy13, scoville15} and others suggesting it is not required \citep[e.g.,][]{reddy15, zeimann15, salmon16}. As can be gathered, there is no consensus regarding the importance of the 2175\ang feature in galaxy attenuation curves, likely owing to large differences in the galaxy samples considered and the methodologies employed for characterizing the dust attenuation. 

Treatments for flexible 2175\ang features in dust attenuation curves are now implemented in several SED modeling codes (e.g., \texttt{Prospector}, \citealt{leja17}; \texttt{CIGALE}, \citealt{boquien19}), and allows for analysis of this feature in much larger samples of galaxies than possible from direct spectroscopy \citep[e.g.,][]{calzetti94, noll09a}. In this paper, we introduce a similar prescription for including a flexible 2175\ang feature in the Multiwavelength Analysis of Galaxy Physical Properties \citep[\magphys;][]{daCunha08, daCunha15} code to explore the strength of the 2175\ang feature and the shape of dust attenuation curves for SFGs at $0.1< z\lesssim3$. This paper is organized as follows: Section~\ref{magphys_method} describes the details of the updated version of \magphys\ used in this study\footnote{Several versions of the \magphys\ code, including the one used in this work, are publicly available online at \url{www.iap.fr/magphys/}.}, Section~\ref{data} presents the observational data, Section~\ref{results} shows our results on the range of observed 2175\ang feature strengths and the derived attenuation curves, and Section~\ref{conclusion} summarizes our main conclusions. Throughout this work we adopt a $\Lambda$-CDM cosmological model, $H_0=70$~km/s/Mpc, $\Omega_M=0.3$, $\Omega_{\Lambda}=0.7$.

\section{Revised \texttt{MAGPHYS} high-$z$ Description}\label{magphys_method} 
\magphys\ \citep{daCunha08, daCunha15} is a SED-fitting code designed to self-consistently determine galaxy properties based on an energy-balance approach using rest-frame UV through radio photometry in a Bayesian formalism. We refer readers to these release papers for details. In brief, \magphys\ uses the stellar population models of \citet{bruzual&charlot03}, assumes a \citet{chabrier03} initial mass function (IMF), and the dust model of \citet{charlot&fall00} for which the interstellar dust is distributed into two components, one associated with star-forming regions (stellar birth clouds) and the other with the diffuse interstellar medium (ISM). 

For this study, we make use of the `high-$z$' version of \magphys\ \citep{daCunha15}, but also introduce an additional component in the attenuation curve for the diffuse ISM to characterize additional attenuation due to a 2175\ang feature. We also alter the prescription for intergalactic medium (IGM) absorption in the ultraviolet from \citet{madau95} to the more recent IGM prescription of \citet{inoue14}. We note that the updated attenuation and IGM prescriptions are identical to those adopted in \magphysz\ \citep{battisti19}. The star formation history (SFH) treatment is unchanged, rising linearly at early ages and then declining exponentially,
with random bursts of star formation superimposed onto the continuous SFH. The dust emission is also unchanged and modeled using templates based on four components: 1) polycyclic aromatic hydrocarbons (PAHs); 2) mid-infrared continuum from hot dust; 3) warm dust in thermal equilibrium; and 4) cold dust in thermal equilibrium \citep{daCunha08}. The details of the updated attenuation models are summarized below. 

The stellar emission of each model is calculated such that the luminosity per unit wavelength emerging at time $t$ from a model galaxy is expressed as:
\begin{equation}
L_{\lambda,\mathrm{obs}}(t)=\int_0^t dt^\prime \Psi(t-t^\prime)\,l^\mathrm{SSP}_\lambda(t^\prime,Z) \,\exp[-\hat\tau_\lambda(t^\prime)] \,,
\label{eq:llambda}
\end{equation}
where $l^\mathrm{SSP}_\lambda(t^\prime,Z)$ is the luminosity emitted per unit wavelength per unit mass by a simple stellar population (SSP) of age $t^\prime$ and metallicity $Z$, $\Psi(t-t^\prime)$ is the star formation rate evolution with time (i.e. the SFH), and $\hat\tau_\lambda(t^\prime)$ is the `effective' absorption optical depth seen by stars of age $t^\prime$\footnote{This relates to the `effective' attenuation through $\hat{A}_\lambda$=1.086$\hat{\tau}_\lambda$.}.
The behavior of $\hat{\tau}_\lambda$ is given by:
\begin{equation}\label{eq:tau_definition}
\hat{\tau}_\lambda(t^\prime)=
\begin{cases}
\hat{\tau}_\lambda^{\,\mathrm{BC}}+\hat{\tau}_\lambda^{\,\mathrm{ISM}} \, & \text{for $t^\prime\leq t_\mathrm{BC}$,} \\
\hat{\tau}_\lambda^{\,\mathrm{ISM}}\, & \text{for $t^\prime> t_\mathrm{BC}$}\,,
\end{cases}
\end{equation}
where $\hat{\tau}_\lambda^{\,\mathrm{BC}}$ is the effective attenuation optical depth of dust in stellar birth clouds, $\hat{\tau}_\lambda^{\,\mathrm{ISM}}$ is the effective attenuation optical depth in the diffuse ISM, and $t_\mathrm{BC}\simeq10^7$~yr \citep{charlot&fall00}. 

For the purpose of this work, it is very important to distinguish between the `effective' optical depth, \tauv, and the `total effective' optical depth, $\tau_\lambda$, where the latter represents the time integration of both the SFH and the effective optical depth in equation~(\ref{eq:llambda}), 
\begin{equation}\label{eq:total_attenuation}
L_{\lambda,\mathrm{obs}}(t)=L_{\lambda,0}(t)e^{-\tau_\lambda}=L_{\lambda,0}(t)10^{-0.4A_\lambda} \,,
\end{equation}
where $L_{\lambda,0}$ is the intrinsic (unreddened) integrated stellar population luminosity per unit wavelength at time $t$, $A_\lambda$ is the total attenuation, and for simplicity we assume a foreground screen geometry. The dust attenuation curves and related quantities derived in Section~\ref{results} are representative of the time-integrated or `total effective' attenuation (e.g., $\tau_\lambda$, $A_\lambda$).

We introduce an additional component into diffuse ISM attenuation curve to account for dust attenuation from a 2175\ang feature. We tested including this feature in only the birth cloud dust component and found that this is insufficient to match observations because the stellar population associated with the diffuse ISM component tends to dominate the observed SED in (massive) galaxies, primarily due to the short birth cloud lifetime and the higher optical depths toward the birth cloud regions. For a majority of the SFGs examined in this work, our models predict that the attenuated SED from the stellar population associated with the birth cloud component typically contributes a very small fraction ($\lesssim$few percent) to the total observed SED in the 2175\ang region, despite the intrinsic (i.e., unattenuated) birth cloud component typically being more luminous at UV wavelengths. As a result, the models are unable to reproduce the observed 2175\ang feature when only including it in the birth cloud component. The lower impact of the birth cloud component on the resulting attenuated SED curve shapes for very dusty galaxies has also been found using \texttt{CIGALE} \citep[e.g.,][]{loFaro17, buat18}. However, it is important to note that the role of the birth cloud component likely changes in the regime of low-mass, high specific-SFR (sSFR=SFR/$M_*$) galaxies. As a final point, it has also been suggested that the dust grains responsible for the feature may be easily destroyed by UV photons in star-forming regions \citep{gordon97, gordon03, fischera&dopita11}. This is supported by the complete absence of this feature in the attenuation curve of local starburst galaxies \citep{calzetti94} and also for its weakened appearance in the LMC2 Supershell region (near 30 Dor) relative to the rest of the LMC \citep{gordon03}. 

The choice of adding a 2175\ang feature to the diffuse ISM attenuation curve in our models is supported from observations of star-forming disk galaxies where the strength of the 2175\ang feature appears dependent on galaxy inclination \citep{wild11, kriek&conroy13, battisti17b}, being strongest for edge-on disk galaxies. This inclination dependence can be explained if the UV emission contributed by star-forming regions, some of which is spatially coincident with the birth cloud dust component (small galaxy covering fraction), experience increased attenuation by the diffuse ISM (high galaxy covering fraction) in higher inclination geometries. We also explored including the 2175\ang feature in both dust components (using equal strength for simplicity) and find that the results do not change significantly from when it is included in only the diffuse component. As a result, we choose to to only include it in the diffuse ISM component.

We make the simplistic assumption that the 2175\ang absorption feature follows a behavior similar to that of the MW extinction curve. The MW 2175\ang feature is well characterized using a Lorentzian-like Drude profile \citep[e.g.,][]{fitzpatrick&massa07}:
\begin{equation}
D(E_b,\lambda) = \frac{E_b(\lambda\,\Delta\lambda)^2}{(\lambda^2-\lambda_0^2)^2+(\lambda\,\Delta\lambda)^2} \,,
\end{equation}
where $\lambda_0$ is the central wavelength of the feature, $\Delta\lambda$ is its FWHM, and $E_b$ is an amplitude constant that defines the bump strength. The average MW extinction curve has values of $\lambda_0=2175.8$\AA, $\Delta\lambda=470$\AA, and $E_b=3.3$ \citep{fitzpatrick99}. The $E_b$ value stated here for the MW refers to the amplitude of the Drude profile in terms of the total-to-selective extinction curve $k_\lambda\equiv A_\lambda/E(B-V)$. However for our purposes, we introduce the Drude profile into attenuation curves defined in terms of normalized optical depth, $\tau_\lambda/\tau_V=(\lambda/5500\ang)^{-n_\mathrm{CF}}$ \citep{charlot&fall00}, and as a result the bump amplitude term takes on a slightly modified meaning, which we denote as $E_b'$ to avoid confusion. The relationship between the two versions of the bump strength is the following: 
\begin{equation}
E_b' = E_b/R_V^{\,\mathrm{ISM}} \,,
\end{equation}
where $R_V^{\,\mathrm{ISM}}$ is the total-to-selective attenuation in the $V$ band from the diffuse ISM (note that $R_V=k_V$). For comparison to dust curves that do not utilize two components (birth cloud and ISM), one can simply adopt $R_V^{\,\mathrm{ISM}}=R_V$. Thus, the MW extinction curve, which has an average value of $R_V=3.1$, has $E_b'=1.06$ \citep{fitzpatrick99}. 

\begin{figure}
\begin{center}
\includegraphics[width=0.46\textwidth]{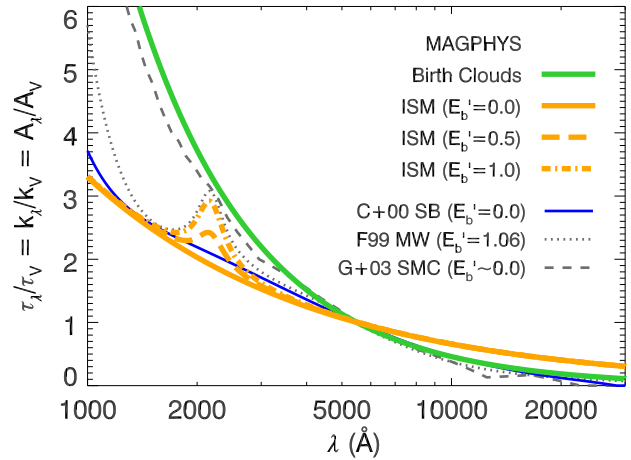}
\end{center}
\vspace{-0.3cm}
\caption{Comparison between the attenuation curves adopted for the birth cloud (green line) and ISM components with differing values of bump strength $E_b'$ (orange lines). The total effective attenuation curve inferred from \magphys\ is a (SFH-dependent) combination of the birth cloud and ISM curves as defined by eq.~(\ref{eq:llambda}) and (\ref{eq:tau_definition}). The starburst attenuation curve \citep{calzetti00} and the average extinction curves of the MW \citep{fitzpatrick99} and SMC \citep{gordon03} are also shown for reference.
\label{fig:att_curve_comparison}}
\end{figure}

The attenuation curve of the birth clouds remains unchanged from previous versions of \magphys, 
\begin{equation}
\hat{\tau}_\lambda^{\,\mathrm{BC}}=(1-\mu)\,\hat\tau_V\,(\lambda/5500\ang)^{-1.3} \,,
\end{equation}
where \tauv\ is the effective $V$-band optical depth seen by stars younger than $t_\mathrm{BC}$ in the birth clouds and $\mu$ is the fraction of \tauv\ seen by stars older than $t_\mathrm{BC}$ (i.e., stars in the diffuse ISM component, $\mu=\hat\tau_V^{\,\mathrm{ISM}}/(\hat\tau_V^{\,\mathrm{BC}}+\hat\tau_V^{\,\mathrm{ISM}})$). We utilize the same $\mu$ prior as \magphys\ high-$z$, which is a Gaussian distribution centered at 0.25 with a standard deviation of 0.10, with values below 0.1 having zero probability. We introduce a Drude profile into the attenuation curve characterizing the diffuse ISM,
\begin{equation}\label{eq:tau_ISM}
\hat{\tau}_\lambda^{\,\mathrm{ISM}}=\mu\,\hat\tau_V\,[(\lambda/5500\ang)^{-0.7}+D(E_b',\lambda)] \,.
\end{equation}
The central wavelength and FWHM of the 2175\ang feature are fixed to the MW value and only the amplitude defining the bump strength, $E_b'$, is free to vary within the code. A demonstration of these curves is shown in Figure~\ref{fig:att_curve_comparison}, where the starburst attenuation curve \citep{calzetti00} and MW extinction curve \citep{fitzpatrick99} are also shown for reference. The assumption of a fixed central wavelength and FWHM may not be accurate for galaxy attenuation curves \citep[e.g.,][]{noll09a}, but we find this is sufficient for the models to reproduce a the majority of the intermediate/broadband photometry considered in our analysis. Furthermore, due to the width of the available filters and their wavelength spacing, the value of central wavelength and the feature width will not be well constrained. The prior distribution that is adopted for $E_b'$ in our updated version of \magphys\ (and also for \magphysz, see \citealt{battisti19}) is presented in Section~\ref{EB_prior}. An important factor to consider when making bump strength comparisons in terms of $E_b$ is that the total-to-selective attenuation curves ($k_\lambda$) inferred from \magphys\ have varying values of $R_V$ that are dependent on the values of $\mu$, \tauv , and the parameters defining the SFH.

\section{Data and Measurements}\label{data}
To reliably characterize dust attenuation, we select galaxies with secure spectroscopic redshifts and mid-IR and/or far-IR data to take full advantage of the energy-balance capabilities of \magphys\ and mitigate the stellar age-dust reddening degeneracy. This degeneracy refers to the similarity in color that a young, dusty stellar population can can have to that of an old, dust-free population \citep[e.g., ][]{witt92,gordon97}. To accurately characterize the shape of the attenuation curve it is also desirable to use samples with extensive multiwavelength photometric data that spans the full range from UV to far-IR wavelengths. We utilize the COSMOS field \citep{scoville07, capak07} because it meets all of these requirements. In particular, the 12 intermediate-band optical data that are available in the COSMOS field from the Subaru telescope \citep{taniguchi07, taniguchi15} are ideal for detecting the presence and strength of the 2175\ang feature at redshifts of $1\lesssim z\lesssim2.8$. We use two photometric+spectroscopic catalogs of galaxies in the COSMOS field that are optimized for different redshift ranges: the first are part of the Galaxy And Mass Assembly (GAMA) survey and the second are produced by the COSMOS team. The galaxy selection method is briefly summarized below and is more extensively outlined in \citet{battisti19}. We note that the IR selection criteria can bias us toward more massive and/or dusty galaxies, on average, relative to a non-IR selected sample, especially at higher redshifts.

\subsection{GAMA - G10 Sample}
The GAMA survey compiled a highly complete multiwavelength photometric \citep{driver11} and spectroscopic sample of galaxies \citep{baldry10, robotham10, hopkins13} covering 280 deg$^2$ to a main survey limit of $r<19.8$ mag that included a subset of the COSMOS field (denoted as G10), although the spectroscopy for this field comes from the zCOSMOS survey \citep{lilly07, lilly09}. The photometric catalog, described in \citet{andrews17}, contains data from \galex, Canada-France-Hawaii telescope (CFHT), Subaru, UltraVista, \spitzer, and \herschel\ that span the full UV to sub-mm wavelength range. The photometry is aperture-matched utilizing the Lambda-Adaptive Multi-Band Deblending Algorithm in R \citep[\texttt{LAMBDAR};][]{wright16} algorithm. The catalog photometry is already corrected for Galactic extinction for all bands from FUV to $K_s$ using $E(B-V)$ values from the \citet{schlegel98} MW dust maps \citet[][see Table 3]{andrews17}. The spectroscopic catalog is described in \citet{davies15} and we utilize the recommended selection criteria of $\mathtt{Z\_BEST}>0.0001$,  $\mathtt{Z\_USE}<3$, and  $\mathtt{STAR\_GALAXY\_CLASS}=0$. This provides a parent sample of 20,364 objects. We further refine this sample to only consider galaxies at $z>0.1$ with $S/N>3$ in any \spitzer/MIPS (24 and 70\micron), \herschel/PACS (100 and 160\micron), or \herschel/SPIRE (250, 350, and 500\micron) band (i.e., requiring a detection in at least one far-IR filter), which leaves 4567 galaxies at $z\lesssim1.6$.
 
\subsection{COSMOS2015 \& Super-Deblended Sample}
To extend the redshift range, we also utilize the latest COSMOS master spectroscopic catalog (curated by M. Salvato for internal use within the COSMOS collaboration), together with the COSMOS2015 \citep{laigle16} and ``Super-deblended'' \citep{jin18} photometric catalogs. The COSMOS2015 catalog provides photometry for the UV through NIR wavelength range from \galex, CFHT, Subaru, UltraVista, and \spitzer/IRAC. The COSMOS2015 photometry is corrected for Galactic extinction from NUV to $K_s$ using the provided $E(B-V)$ and adopting the same extinction coefficients as G10 for consistency \citep{andrews17}. We adopt total flux values (3\arcsec + aperture correction) using the prescription described in \citet{laigle16}, as these are better suited for combining with the total flux measurements of IR data in the Super-deblended catalog. The Super-deblended catalog extends the wavelength coverage for the mid-IR through radio wavelength range from the \spitzer/MIPS, \herschel, SCUBA2, AzTEC, MAMBO, and Very Large Array (VLA). These catalogs have been cross-matched with X-ray sources from the Chandra catalog \citep{elvis09, civano12, civano16, marchesi16} and the XMM/Newton Wide-Field Survey \citep{hasinger07, cappelluti09}. For brevity, we will refer to the combined COSMOS2015+Super-deblended catalogs as ``C15+SD catalog" throughout the remainder of the paper. 

For our parent sample of spectroscopic galaxies, we utilize cases with robust spectroscopic redshifts  \citep[Quality flag $\mathtt{Q_f}=3$ or 4, as defined in zCOSMOS;][]{lilly09}. The COSMOS master catalog contains duplicate observations and these are remedied by giving preference to higher quality data and newer observations \citep[see][]{battisti19}. After duplicate removal, we are left with a sample of 34,785 galaxies. Further refining this sample to only consider galaxies at $z_\mathrm{spec}>1.0$ with $S/N>3$ in any band from \spitzer/MIPS (24\micron), \herschel/PACS (100 and 160\micron), \herschel/SPIRE (250, 350, and 500\micron), SCUBA (850\micron), AZTEC (1.1mm), or MAMBO (1.2mm), leaves 2133 galaxies at $1\leq z<6$. For cases where the same galaxy is present in both the G10 and C15+SD catalogs, we adopt the G10 photometry. This removes 156 galaxies from the C15+SD catalog, leaving a sample of 1977.

\subsection{AGN Identification and Removal}\label{AGN_removal}
We identify and remove active galactic nuclei (AGN) from the analysis because current versions of \magphys\ are intended only for purely star forming galaxies. AGN are identified using the following techniques: 1) the \spitzer/IRAC color selections of \citet{donley12}; 2) the \spitzer-\herschel\ color selections of \citet{kirkpatrick13}; 3) the radio-NIR color selection of \citet{seymour08}; and 4) sources with any X-ray detection. We note that for the color selection criteria each photometric band used is required to have signal-to-noise ratio of $S/N>3$. These methods remove 651 AGN (14.3\%) from the G10 sample and 110 AGN (5.6\%) from the C15+SD sample. In general, the various AGN selection methods do not overlap ($<$10\% AGN identified with multiple diagnostics), but they are complimentary in providing a more complete method for AGN identification. AGN selected via IR colors are most likely to have poor \magphys\ fits \citep[large $\chi_\mathrm{red}^2$; see][Appendix A]{battisti19}. This leaves a final sample of 3916 and 1867 galaxies in the G10 and C15+SD catalogs, respectively, for our analysis, primarily spanning $0.1< z\lesssim3$ ($<$1\% lie at $z>3$). 



\subsection{Galaxy Inclination Measurements}
The geometric distribution of dust and stars is expected to have a significant effect on the resulting dust attenuation. Local, edge-on galaxies experience up to 1~mag of additional attenuation in the $B$- or $g$-bands relative to face-on galaxies \citep[e.g.,][]{disney89, giovanelli94, masters03, masters10, driver07, unterborn&ryden08, maller09, yip10}. Radiative transfer simulations predict that attenuation curves should become shallower (or grayer) at higher inclinations due to increasing optical depth, which is a result of differential optical depth effects \citep[bluer light arises from shallower physical depths than redder light; e.g.,][]{calzetti01, pierini04,chevallard13, seon&draine16}. This graying effect is observed in the attenuation curves of local spiral galaxies \citep[e.g.,][]{wild11, battisti17b}. Therefore, we explore if any trends are evident in the behavior of the dust attenuation curve or 2175\ang bump strength with galaxy inclination.

To provide the highest spatial resolution, we use the \hst\ ACS/WFC catalog available on the NASA/IPAC website\footnote{\url{https://irsa.ipac.caltech.edu/data/COSMOS/}} \citep{leauthaud07, koekemoer07}. The values in this catalog are based on SExtractor \citep[version 2.4.3][]{bertin&arnouts96}. The axial ratio (a proxy for inclination) for each galaxy, in both samples used, is determined using the \texttt{a\_world} and \texttt{b\_world} parameters that define the profile RMS along major and minor axis, respectively. When making comparisons using axial ratios, we restrict the sample to sources with minor axes larger than 3.6 pixels \citep[0.108",][]{leauthaud07}. We explored using different size cuts but found that this does not significantly affect the results.

\section{Application of Revised \texttt{MAGPHYS} code to COSMOS}\label{results}
\subsection{Evidence for the 2175\texorpdfstring{\ang}{Ang} Feature and Constructing an Appropriate Prior for \texorpdfstring{$E_\mathrm{\lowercase{b}}'$}{Eb}}\label{EB_prior}
\begin{figure*}
\begin{center}
$\begin{array}{c}
\includegraphics[width=0.85\textwidth,clip=true]{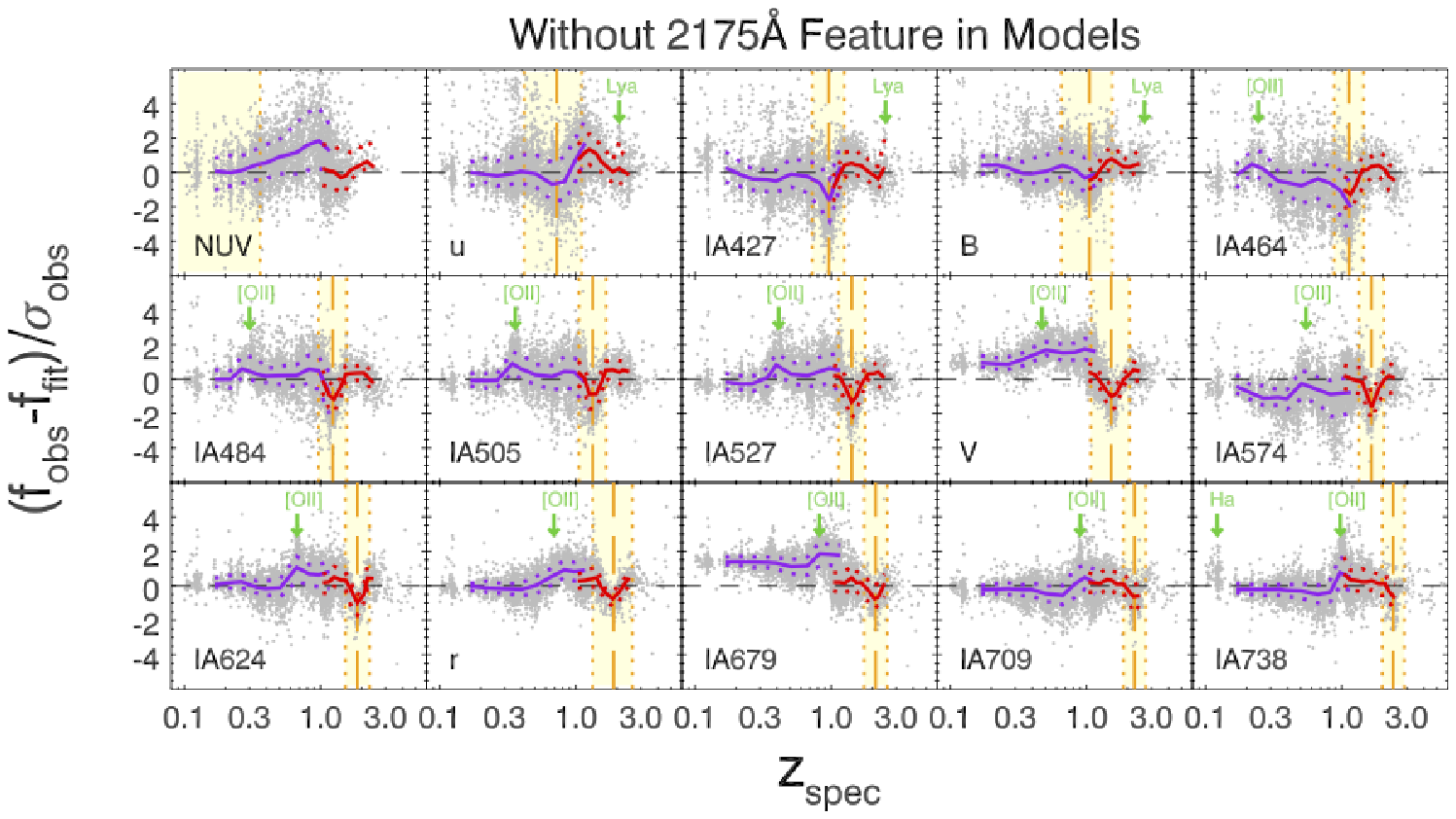} \\
\includegraphics[width=0.85\textwidth,clip=true]{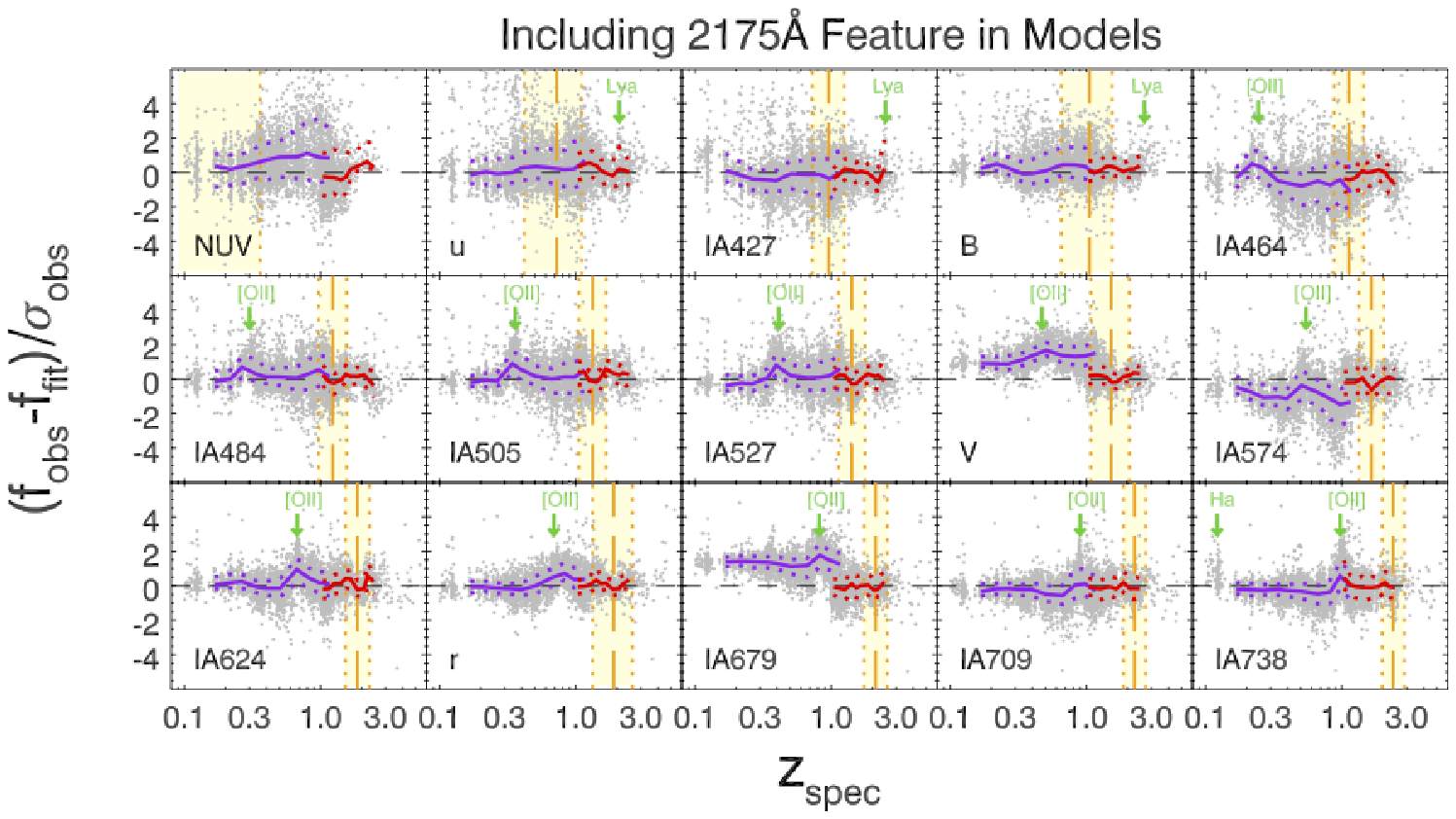} \\
\end{array}$
\end{center}
\vspace{-5mm}
\caption{\textit{Top:} Residuals between a subset of the observed photometry and the predicted values from default \magphys\ models when no 2175\ang feature is included as a function of redshift for all IR-detected SFGs in our sample (gray circles). The median value and 16th and 84th percentiles are shown by the colored (G10=magenta, C15+SD=red) solid and dotted lines, respectively. The redshift where each filter corresponds to rest-frame 2175\ang is the denoted by the dashed orange line and a representative range of influence by the dotted orange line. The panels show clear evidence of 2175\ang absorption as a deficit between the observed and model flux, $(f_\mathrm{obs}-f_\mathrm{fit})/\sigma_\mathrm{obs}$, at most redshifts. The redshift where strong emission lines fall into a filter are also shown and results in a separate deviation due to their exclusion in the models. Residual offsets are apparent in some bands between the two samples and are due to different adopted zeropoints between the catalogs (e.g., $V$-band) and/or different filters used in the fits (e.g., G10 includes FUV). \textit{Bottom:} The photometric residuals when using \magphys\ models that include a 2175\ang feature. The panels show a dramatic improvement in the region where 2175\ang absorption occurs. 
\label{fig:2175_residual}}
\end{figure*}

During early tests of the \magphys\ code on our sample of star-forming galaxies, it became evident that including a 2175\ang absorption feature is required for the models to properly fit the SED and avoid large residuals in the region surrounding 2175\AA. The necessity of a feature in the models is easily apparent when we examine the residuals of the observed photometry and the predicted values from default \magphys\ models, as shown in Figure~\ref{fig:2175_residual}, \textit{Top} for the G10 and C15+SD sources in our primary sample. The region where the 2175\ang feature is expected to lie is denoted and clear deviations in this region are evident, especially for the intermediate Subaru bands (those starting with ``IA'').

To establish an appropriate prior distribution for the 2175\ang bump in \magphys, we utilize a subset of our full sample and only consider galaxies where at least two bands are detected ($S/N>3$) in the 2175\ang feature region ($\mathrm{FWHM}=470$\AA), defined as $1940\mathrm{\ang}<(\lambda_\mathrm{filt}\pm\mathrm{FWHM}_\mathrm{filt}/2)/(1+z)<2410\mathrm{\ang}$. This criteria limits us to $z\gtrsim0.6$ for the G10 sample ($N=1571$) but retains most of the C15+SD sample ($N=1817$). As a reminder, the bump feature is added only to the diffuse ISM dust curve in the models (see Section~\ref{magphys_method} for more details). We make the simple initial assumption that any bump strength in the range $0\le E_b'\le 3.3$ is equally likely to occur (i.e., a flat prior distribution from $\sim0-3\times$MW strength). We repeated the fitting sequence using the posterior $E_b'$ values of the previous run as a new prior to reduce the impact of the initial choice on the fitting results. This process is repeated until the results converge and this is achieved after 4 iterations. 


\begin{figure}
\begin{center}
\includegraphics[width=0.45\textwidth,clip=true]{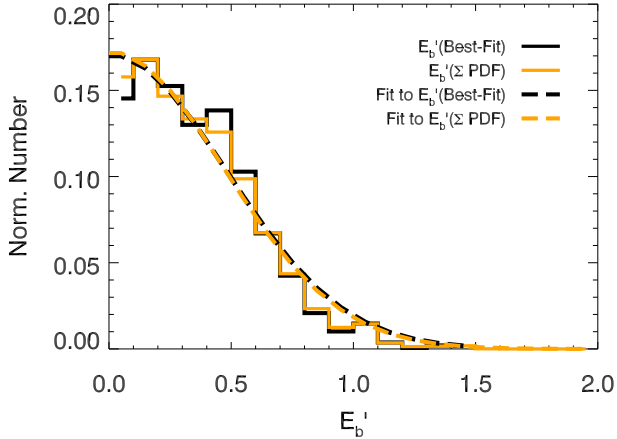}
\end{center}
\vspace{-0.4cm}
\caption{Histogram of $E_b'$ values for the combined GAMA/G10 and C15+SD sources for which at least 2 bands of $S/N>3$ sample the bump region. Both the best-fit values (black solid line) and the summed posterior PDFs (orange solid line), where the latter includes full likelihood distributions, show a similar trend of rapidly decreasing likelihood at higher values of $E_b'$. We adopt a Gaussian fit of the summed-PDFs (black dashed line) in the range of $0\le E_b'\le 1.5$ as the input prior for \magphys.
\label{fig:Eb_hist}}
\end{figure}

The best-fit and posterior probability distribution function (PDF) of $E_b'$ values for both samples from the final iteration are shown in Figure~\ref{fig:Eb_hist} and demonstrates a rapid decline in probability with increasing values of $E_b'$. We do not distinguish between the sample populations because both appear to have a similar $E_b'$ distribution. In addition, there does not appear to be a substantial change in the $E_b'$ distribution or its dispersion as a function of redshift for this sample of SFGs. We examine the behavior of $E_b'$ as a function of various physical properties in Section~\ref{dust_curves}. The trend we find favoring low bump strengths (recall MW strength is $E_b'=1.06$) is not surprising because both geometric and scattering effects act to suppress the apparent strength of a bump feature in attenuation curves relative to extinction curves \citep[e.g.,][]{natta&panagia84, calzetti94}. We fit each distribution using a Gaussian function and the results are shown in Figure~\ref{fig:Eb_hist}. For the fitting procedure we mirror the $E_b'$ distribution on the negative side so that the function is centered at zero. During the iteration process, the probability of strong bump cases were quickly found to be very low and we only utilize the range of $0\le E_b'\le 1.5$ for the final adopted prior (i.e., probability of $E_b'>1.5$ is zero) for the updated \magphys\ high-$z$. The median value of $E_b'$ and the 16th to 84th percentile range are $0.31^{+0.34}_{-0.22}$. These values are in good agreement with findings for SFGs at both low-$z$ \citep[e.g.,][]{salim18} and intermediate-$z$ \citep[e.g.,][]{buat12}. The improvement in the SED fits afforded by the inclusion of a 2175\ang feature, using the new prior, is demonstrable by looking at the residuals of the photometric and model data as a function of redshift, as shown in Figure~\ref{fig:2175_residual}, \textit{Bottom}. 

The change in $\chi_\mathrm{red}^2$ values before and after the inclusion of the 2175\ang feature is also a good indicator of the significance of the feature. We note that in our case $\chi_\mathrm{red}^2=\chi^2/N_\mathrm{bands}$, where $N_\mathrm{bands}$ is the number of non-zero bands observed, instead of the typical definition of normalizing by the number of degrees of freedom. When comparing the values for the full SED fits, the differences in $\chi_\mathrm{red}^2$ are quite small owing to the few bands that probe the region at a given redshift relative to the total number of bands (median of 31 and 27 bands with $S/N>3$, for G10 and C15+SD samples, respectively). Considering only the subsample utilized in this section (i.e., at least two bands probing the feature), the median $\chi_\mathrm{red}^2$ value changes from 1.50 to 1.36 for G10 without and with the 2175\ang feature included, respectively. This change is 0.71 to 0.56 for the C15+SD sample, respectively, where the lower values in the C15+SD sample relative to G10 are primarily due to the lower $S/N$ of higher redshift data. Perhaps a better indicator is the median $\chi_\mathrm{red}^2$ value for only the filters probing the 2175\ang feature. For this subset of filters, the median $\chi_\mathrm{red}^2$ value changes from 1.33 to 0.77 for G10 without and with the 2175\ang feature included, respectively. This change is 0.82 to 0.37 for the C15+SD sample, respectively. For both samples, the median $\chi_\mathrm{red}^2$ value is considerably lower for the filters probing the 2175\ang feature when it is included in the models. 

Another method for quantifying the improvement is to examine the Bayesian Information Criterion (BIC) \citep[e.g.,][]{liddle07}. Under the simplified assumption that the posterior distributions are Gaussian, the BIC can be calculated as:
\begin{equation}\label{eq:BIC}
\mathrm{BIC}=\chi^2 +k~\ln N \,,
\end{equation}
where $\chi^2$ corresponds to the minimum value (maximum likelihood), $k$ is the number of free parameters in the models, and $N$ is the number of data points used in the fit. For standard \magphys\ (and \magphys\ high-$z$), the number of free parameters in the models is $k=13$, and adding in a parameter for the 2175\ang feature increases this to $k=14$. For the subsample used in this section, we find $\mathrm{BIC}(\mathrm{no~bump})-\mathrm{BIC}(\mathrm{incl.~bump})=1.2$ and 1.5 for the G10 and C15+SD samples, respectively. These values indicate a slight preference towards the fits including the 2175\ang feature, because it has a lower BIC, however the strength of this evidence is very weak and falls under ``not worth more than a bare mention'' category \citep[e.g.,][]{kass&raftery95}. Similar to the case for  $\chi_\mathrm{red}^2$, we suspect the evidence is weakened by the fact that only a few bands probe the feature relative to the total number of bands. Despite the low formal evidence in favor of including the 2175\ang feature in the models, we find that it can play a significant role in estimates of photo-$z$'s for very dusty galaxies and for that reason it is important to account for \citep{battisti19}.


It is worth discussing further the decision to utilize a prior for the $E_b'$ parameter instead of letting it freely vary. The primary motivation is driven by the desire to limit the bump strength from being significantly overestimated when limited data in the rest-frame UV are available. In particular, this has a significant effect in \magphysz\ \citep{battisti19}, where the redshift is a free parameter, because the Lyman break feature (912\AA) can be mimicked by a very strong 2175\ang feature, resulting in incorrect photo-$z$ estimates. The adopted prior favors lower bump strength values while still allowing a relatively wide a range bump strengths ($\sim0-1\times$MW strength).

As mentioned earlier, the IR-detection criteria we utilize tends to select more massive and/or dustier galaxies, on average, relative to a non-IR selected sample. Thus, one may wonder if the $E_b'$ distribution for IR-detected galaxies is well suited to adopt as a prior for all galaxies. For example, based on the different observed strengths of the 2175\ang feature in MW, LMC, and SMC extinction curves, it is possible that the feature is metallicity-dependent. Therefore, it could be the case that IR-detected galaxies are predisposed towards stronger bump features due to higher average metallicities \citep[mass-metallicity relation; e.g.,][]{tremonti04}. We explored the role of the $E_b'$ prior on the fitting results for galaxies without IR-detections, defined as sources not meeting the criteria of $S/N>3$ in a band at $\lambda\geq24\micron$, in the COSMOS field and find they do not show a preference toward lower $E_b'$ values \citep[for details, see][]{battisti19}. However, it is very important to note that the total attenuation of the bump feature scales directly with the amount of reddening ($A_{2175}=E_b\cdot E(B-V)\propto E_b'\cdot A_V$; this is ignoring effects of the filter response), such that the choice of the prior has a decreasing impact on the fitted SED for galaxies with lower reddening. This scaling also implies that constraints on $E_b'$ are poorer for lower $A_V$ galaxies. As a reference, the IR-detected C15+SD sample experiences larger average total attenuation than galaxies without an IR detection ($\bar{A}_V(\textrm{IR-det.})$=0.91, $\bar{A}_V(\textrm{non-IR})$=0.46). As a result of these findings, we believe it is reasonable to adopt the distribution shown in Figure~\ref{fig:Eb_hist} for the general application of \magphys\ on most galaxies.

\subsection{Total Effective Dust Attenuation Curve and Trends with Physical Properties}\label{dust_curves}
Using the updated \magphys\ high-$z$ code, we perform fits to our entire sample of galaxies in order to examine the total effective dust attenuation curves that are inferred. As a reminder, \magphys\ assumes fixed-slopes for the attenuation curve of the two dust components in the models, but the shape of the total effective dust curve can vary due to the different fractional contribution of each component to the total optical depth (depends on $\mu$ and the SFH). For the following analysis, we restrict it to galaxies for which $A_V>0.2$ and $\chi_\mathrm{red}^2<\bar{\chi}_\mathrm{red}^2+4\sigma(\chi_\mathrm{red}^2)$, where $\bar{\chi}_\mathrm{red}^2$ and $\sigma(\chi_\mathrm{red}^2)$ are the mean and dispersion of a Gaussian fit to the lower 90\% of the $\chi_\mathrm{red}^2$ population (determined separately for G10 and C15+SD), that are considered to have more reliably derived attenuation curves. The fits of the $\chi_\mathrm{red}^2$ distribution give $\bar{\chi}_\mathrm{red,G10}^2=0.91$ and $\sigma(\chi_\mathrm{red,G10}^2)=0.43$, and $\bar{\chi}_\mathrm{red,C15+SD}^2=0.51$ and $\sigma(\chi_\mathrm{red,C15+SD}^2)=0.20$. The $A_V$ and $\chi_\mathrm{red}^2$ cuts remove 567 and 531 galaxies (46 in both), respectively, leaving 4731 galaxies for the analysis. The stellar mass of this sample primarily spans $9.5\lesssim \log(M_*/M_\odot)\lesssim 11.3$ (2$\sigma$ range; median of 10.5) and the SFR spans $-0.4\lesssim \log(SFR/(M_\odot \mathrm{yr}^-1))\lesssim 2.0$ (median of 1.0).

\begin{figure*}
\begin{center}
\begin{minipage}{0.6\textwidth}
\includegraphics[width=\textwidth]{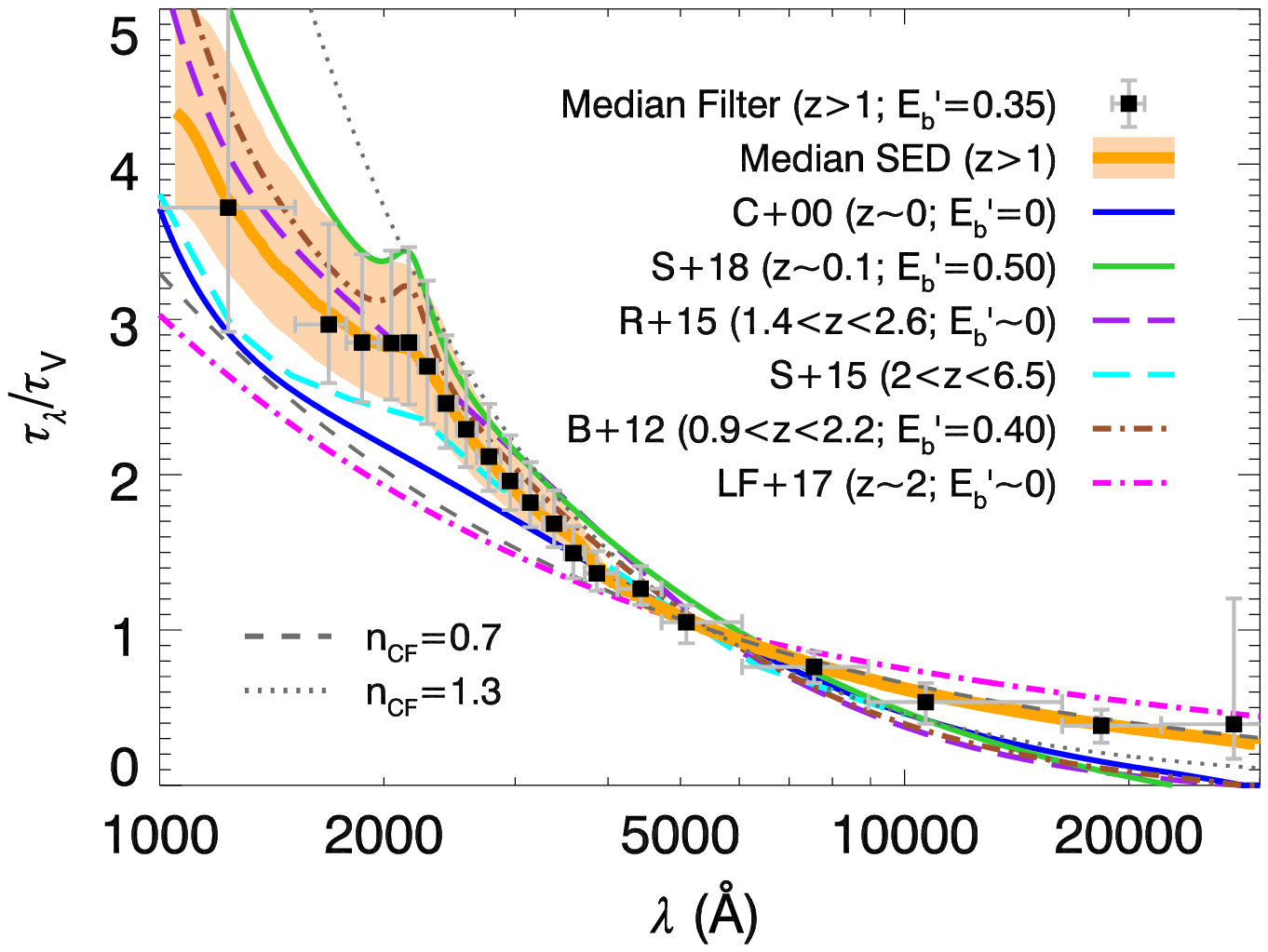}
\end{minipage}
\hspace{1mm}
\begin{minipage}{0.35\textwidth}
\includegraphics[width=\textwidth,clip=true]{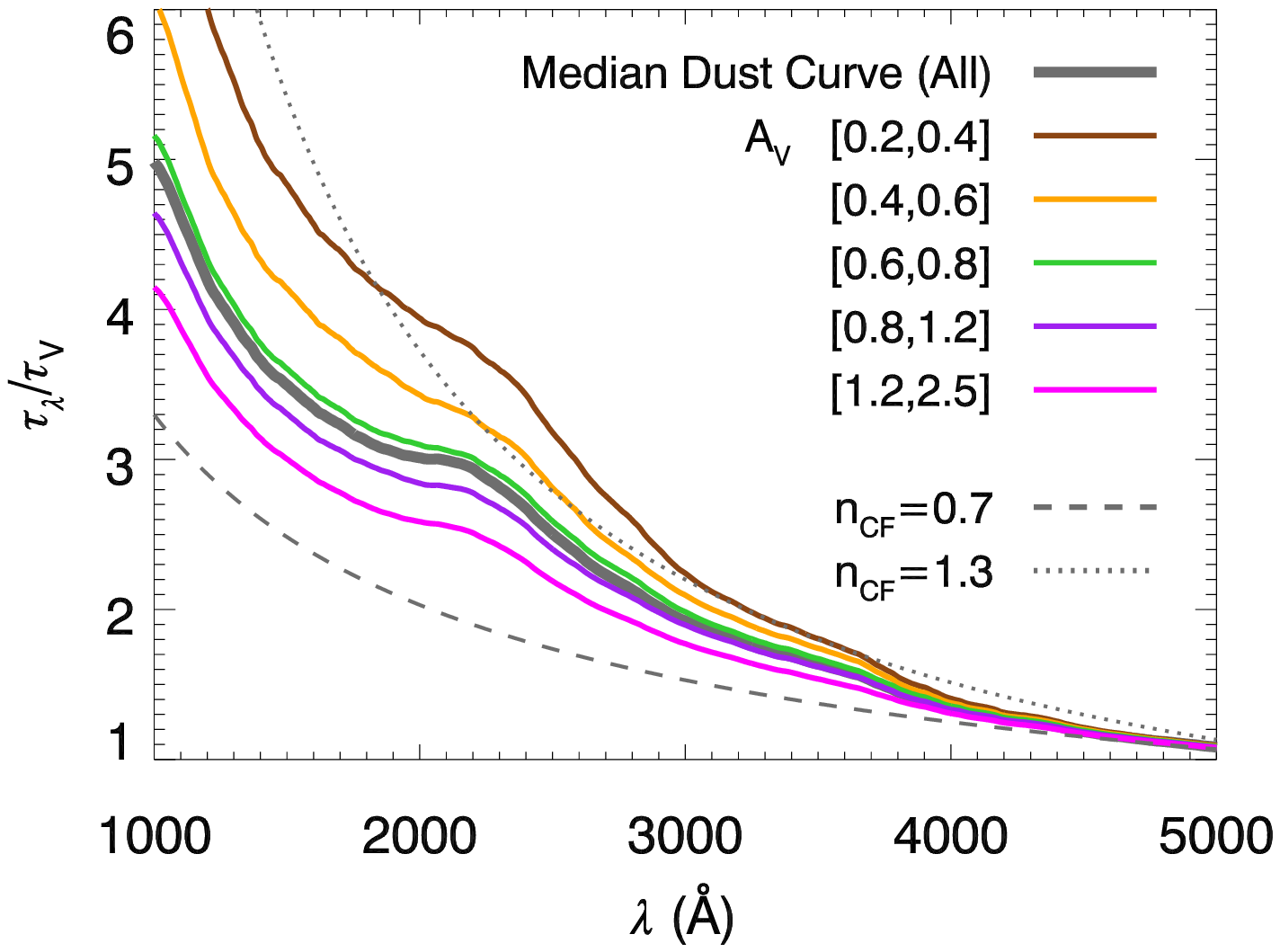} \\
\includegraphics[width=\textwidth]{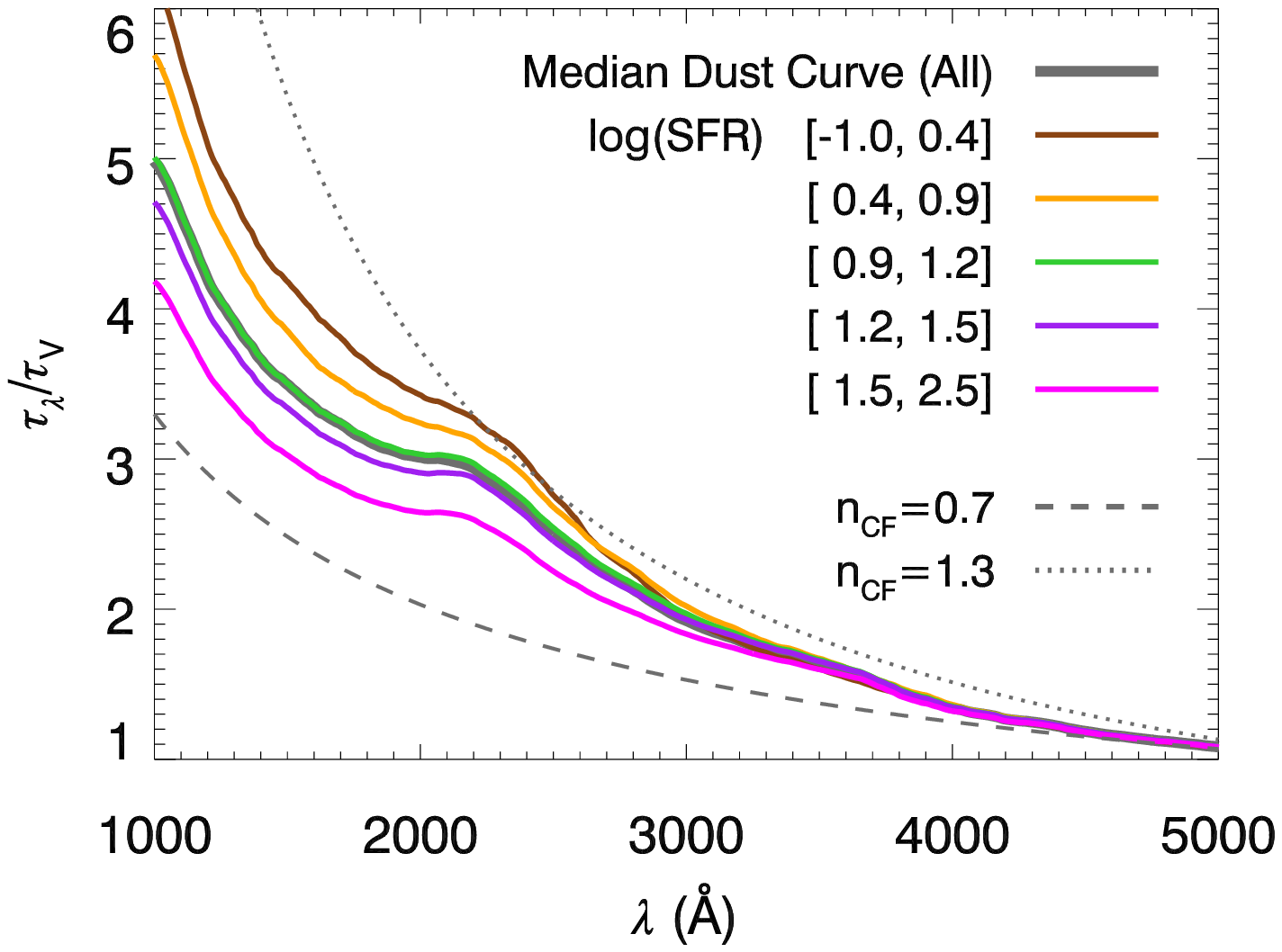}
\end{minipage}
\end{center}
\vspace{-0.3cm}
\caption{\textit{Left:} Average dust attenuation curve, normalized to $V$-band, for galaxies at $1\leq z\lesssim3$ derived from the median value of filters at similar rest-frame wavelength (black squares) and the median of the best-fit SED model (orange line), together with their 1$\sigma$ range, which are consistent. The average attenuation curves of local starbursts \citep[blue line,][]{calzetti00}, local SFGs \citep[green line,][]{salim18}, intermediate-$z$ SFGs (purple line, \citealt{reddy15}; brown dash-dot, \citealt{buat12}), intermediate-$z$ (ultra-)luminous IR galaxies (magenta dash-dot; \citealt{loFaro17}), and high-$z$ SFGs \citep[cyan line,][]{scoville15} are shown. The power-law shape of the diffuse ISM ($n_\mathrm{CF}^{\,\mathrm{ISM}}=0.7$) and birth cloud ($n_\mathrm{CF}^{\,\mathrm{BC}}=1.3$) attenuation components are also shown for reference.  \textit{Right:} Average dust attenuation curve for all  galaxies (gray line) and subdivided by $A_V$ and SFR. For clarity, we do not show the uncertainty or the photometric-based average curve in these panels. The differences between the literature curves in the \textit{left} panel may reflect underlying differences in the physical properties of the galaxy samples used for each study. 
\label{fig:dust_curves}}
\end{figure*}

Our derived attenuation curves are obtained from comparing the best-fit model of the observed SED, $L_{\lambda,\mathrm{obs}}$, relative to the intrinsic stellar population model SED without dust attenuation, $L_{\lambda,0}$, as given by equation~\ref{eq:total_attenuation} but also accounting for the IGM attenuation. As a consistency check, we also determine the attenuation curve inferred from the photometry relative to the predicted intrinsic SED model. The average attenuation curve that is inferred for the $z\geq1$ sample is shown in Figure~\ref{fig:dust_curves}, \textit{Left}, with some other curves in the literature also shown for reference. The photometric attenuation curve is obtained by taking the median value for filters spanning similar regions of rest-frame wavelength in 20 equal-number bins using all filters out to IRAC ch4 (8\micron). We demonstrate the average curve inferred from the $z\geq1$ sample (median of $z=1.2$) because galaxies at $1\lesssim z\lesssim2.8$ have Subaru intermediate band data in the 2175\ang region and the presence of the bump is clearly evident from the median photometric attenuation curve and shows good agreement with the attenuation curve derived from the median best-fit SED models. At UV wavelengths, the slope of our dust curve for intermediate-$z$ SFGs is most similar to the curve found by \citet{reddy15} for galaxies at comparable redshifts, although they do not infer the presence of the bump. Our average attenuation curves, regardless of redshift, have $E_b' \sim 0.3$, which are comparable to median values in \citet[][low-$z$]{salim18}, \citet[][intermediate-$z$]{buat12}, and \citet[][high-$z$]{scoville15}. At $\lambda>5500$\AA, the differences between our curve and the others, with the exception of \citet[][$n_\mathrm{CF}^{\,\mathrm{ISM}}\sim0.48$]{loFaro17} which examined (ultra-)luminous IR galaxies, is attributed to the assumed shape of the diffuse ISM dust curve (dashed gray line; $n_\mathrm{CF}^{\,\mathrm{ISM}}=0.7$). In Appendix~\ref{app_nISMvary}, we explore the impact of allowing $n_\mathrm{CF}^{\,\mathrm{ISM}}$ to vary on the derived attenuation curves.
The average attenuation curve of the entire sample over all redshift is slightly steeper at UV wavelengths than the $z\geq1$ sample (shown in Figure~\ref{fig:dust_curves}, \textit{Right}). As will be discussed below, we attribute changes in the shape of these samples to differences in physical properties.

Next, we explore the behavior of the attenuation curve for subsamples of galaxies with different physical properties. More specifically, we examined the effect of dividing the entire sample, combining both G10 and C15+SD, by stellar mass ($M_*$), star formation rate (SFR), sSFR, total effective $V$-band attenuation ($A_V$), total IR luminosity ($L_\mathrm{dust}$), and inclination via the axial ratio ($b/a$). The strongest trend with the curve slope is found for $A_V$, with weaker trends also seen for SFR, sSFR, and $L_\mathrm{dust}$. We also examined the relationship with $\mu$ (not shown) but do not find any significant correlation because of the additional dependence of the total effective dust curve on the SFH. We show the median curves based on best-fit SEDs in Figure~\ref{fig:dust_curves}, \textit{Right}, for subdivisions of $A_V$ and SFR. We find that the total effective dust attenuation curves becomes shallower for galaxies with increasing $A_V$, SFR, sSFR, or  $L_\mathrm{dust}$, corresponding to being closer in shape to the diffuse ISM dust component. It is important to note that these quantities are not independent, because larger SFRs (or sSFR or $L_\mathrm{dust}$) will tend to occur in systems with larger $A_V$ due to more abundant gas \citep[assuming a dust-to-gas ratio; e.g.,][]{draine07}. 

\begin{figure*}
\begin{center}
\includegraphics[width=0.95\textwidth,clip=true]{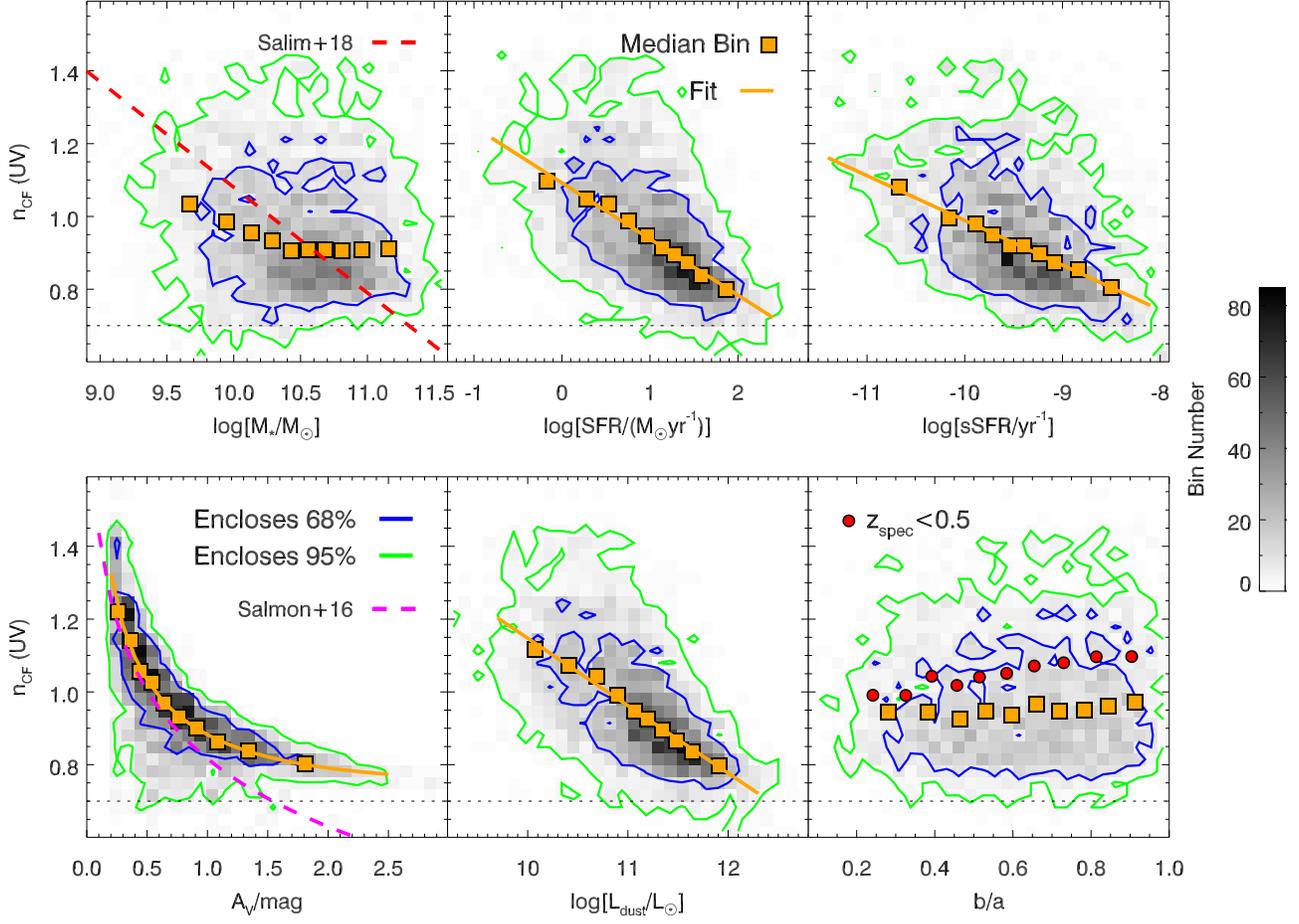}  
\end{center}
\vspace{-0.4cm}
\caption{Distribution of the power-law slopes in the UV region of the total effective dust attenuation curve, $n_\mathrm{CF}\mathrm{(UV)}$ ($\tau_\lambda/\tau_V=(\lambda/5500\ang)^{-n_\mathrm{CF}}$), as a function of $M_*$, SFR, $A_V$, axial ratio ($b/a$). Larger values of $n_\mathrm{CF}\mathrm{(UV)}$ correspond to a steeper attenuation curve. The median value for equal-number bins are indicated by orange boxes. The strongest trends are observed with SFR and $A_V$ (fit shown by solid orange line) and indicate that galaxies with higher SFRs and dust content have shallower total effective dust attenuation curves, with the latter being comparable to the results of \citet{salmon16} (cyan dashed line). The relation with $M_*$ for local galaxies from \citet{salim18} (red dashed line) is shown for reference. Inclination appears to influence the curve slope only for lower redshift galaxies (red circles; see text for details). The dotted black line corresponds to $n_\mathrm{CF}^{\,\mathrm{ISM}}=0.7$ and is a ``lower limit'' of the shallowness of the dust curve slope in our formalism (see text for details). We explore the effect of allowing $n_\mathrm{CF}^{\,\mathrm{ISM}}$ to vary in Appendix~\ref{app_nISMvary}.
\label{fig:nCF_2Dhist}}
\end{figure*}

For quantitative comparison, we determine the slope of the UV portion of the total effective dust curve ($n_\mathrm{CF}\mathrm{(UV)}$; $\tau_\lambda/\tau_V=(\lambda/5500\ang)^{-n_\mathrm{CF}}$) using a power-law fit of regions that avoid the 2175\ang feature (we use $1050\mathrm{\AA}<\lambda<1385\mathrm{\AA}$ and $3000\mathrm{\AA}<\lambda<3700\mathrm{\AA}$). At $\lambda>5500$\ang, the attenuation curves for all subsamples are very similar to the diffuse ISM curve ($n_\mathrm{CF}^{\,\mathrm{ISM}}=0.7$) and are not shown. The similarity of the curves at longer wavelengths is due to our sample consisting primarily of massive galaxies such that the older stellar populations ($t^\prime> 10^7$~yr; associated with the diffuse ISM) provide the dominant contribution to the SED at these wavelengths. Furthermore, it is important to recall that in the \citet{charlot&fall00} formalism the youngest stellar populations experience attenuation from \textit{both} dust components. In a situation of comparable contribution from both components to the total effective optical depth ($\tau_V$), the birth cloud dust will dominate the attenuation at shorter wavelengths and the diffuse dust will dominate at longer wavelengths (the exact wavelength demarcation depends on $\mu$ and the SFH). In Figure~\ref{fig:nCF_2Dhist}, we show the distribution of UV dust curve slopes, $n_\mathrm{CF}\mathrm{(UV)}$, as a function of various galaxy properties. A summary of the Spearman $\rho$ and Kendall $\tau$ rank correlation coefficients\footnote{We do not report the significance of these correlation
coefficients because they are not meaningful for very large sample sizes.} and functional fits, when appropriate, for these parameters are shown in Table~\ref{tab:fit_metrics}.

The trend of the curve slope with $A_V$ is predicted by simulations using radiative transfer, which show that effects of differential optical depth result in attenuation curves becoming flatter with increasing total optical depths \citep[e.g.,][]{pierini04, chevallard13, seon&draine16, narayanan18}. It is important to note that, in these simulations, the flattening occurs assuming the same intrinsic extinction curve (i.e., it is purely due to geometric and scattering effects). These trends are also found in other observational studies \citep[e.g.,][]{salmon16, salim18} and are in good agreement with our findings. For comparison to other works that utilize a modified starburst attenuation curve, $k(\lambda)\propto k_\mathrm{SB}(\lambda) (\lambda/5500\ang)^\delta$, where $k_\mathrm{SB}$ is from \citet{calzetti00} and $\delta$ modifies the slope, the $\delta$ can be related to the power-law slope in our formalism \citep{charlot&fall00} through $\delta\simeq0.7-n_\mathrm{CF}$. We show the relation between the dust curve slope and $A_V$ from \citet[][using their eq~(8) and $A_V=E(B-V)/4.05$ in their formalism]{salmon16} in Figure~\ref{fig:nCF_2Dhist}. The different behavior between our results at higher $A_V$ is attributed to the ``lower limit'' of $n_\mathrm{CF}^{\,\mathrm{ISM}}=0.7$ in the \citet{charlot&fall00} formalism. We explore the effect of allowing $n_\mathrm{CF}^{\,\mathrm{ISM}}$ to vary in Appendix~\ref{app_nISMvary} and find that the highest $A_V$ sources show a preference toward models with values of $n_\mathrm{CF}^{\,\mathrm{ISM}}<0.7$. This highlights that the trends shown are dependent on our assumed priors. It is important to note that the power-law indices of the two dust components only represent soft boundaries to the shape of the total effective attenuation curve because these components are applied in a stellar age-dependent manner such that the attenuation is also dependent on the SFH of the galaxy \citep{narayanan18}. This is most apparent for low $A_V$ cases, where the inferred slope of the total effective dust curve can be steeper than the birth-cloud dust component ($n_\mathrm{CF}\mathrm{(UV)}>1.3$). 

It is worth discussing the SFR trend in the context of the \citet{charlot&fall00} formalism. One might intuitively expect that, as the SFR increases, the fractional contribution of recently formed stars associated with the birth cloud component ($t^\prime\leq t_\mathrm{BC}$) would also increase on the observed SED, and that the dust curve should become more similar in shape to the birth cloud component ($n_\mathrm{CF}^{\,\mathrm{BC}}=1.3$). However, the effect just described is actually counter to the observed trend. A closer inspection of the fitting results indicate that while the fractional contribution of the young stellar populations, associated with the birth cloud dust component, to the \textit{intrinsic} SED does increase with increasing SFR relative to the older diffuse stellar population, the young stars become increasingly obscured and end up providing a decreasing fractional contribution to the total \textit{attenuated} SED. Therefore, the total effective attenuation curve primarily reflects the difference between the intrinsic SED and the older, less-dusty stellar population that is influenced by the diffuse dust component. Similar arguments also apply to sSFR and $L_\mathrm{dust}$ because of their direct relation to the SFR.

In local galaxies, it is seen that the attenuation curve slope becomes shallower with increasing galaxy inclination (decreasing axial ratio) \citep[e.g.,][]{wild11, battisti17b, salim18}. This effect can be attributed to the increase in dust column density ($A_V$) as a galaxy becomes more edge-on \citep[e.g.,][]{chevallard13}. Interestingly, we do not see any notable trends in the attenuation curve slope with axial ratio (determined from \hst /ACS) when considering the full galaxy sample over all redshift. However, we do find a weak trend toward shallower slopes at lower axial ratio when we restrict the sample to $z\lesssim0.5$, consistent with local trends found by others. It is possible that this effect is due to galaxies at earlier cosmic time being more clumpy and less settled than local galaxies \citep[e.g.,][]{elmegreen04, elmegreen&elmegreen06, wisnioski15, simons17}. These results do not change when performing different angular size cuts, indicating that this is not a consequence of nearby galaxies being preferentially larger and having more reliable axial ratio measurements than more distant sources. We plan to explore this topic further in a subsequent paper. 

\begin{figure*}
\begin{center}
\includegraphics[width=0.95\textwidth,clip=true]{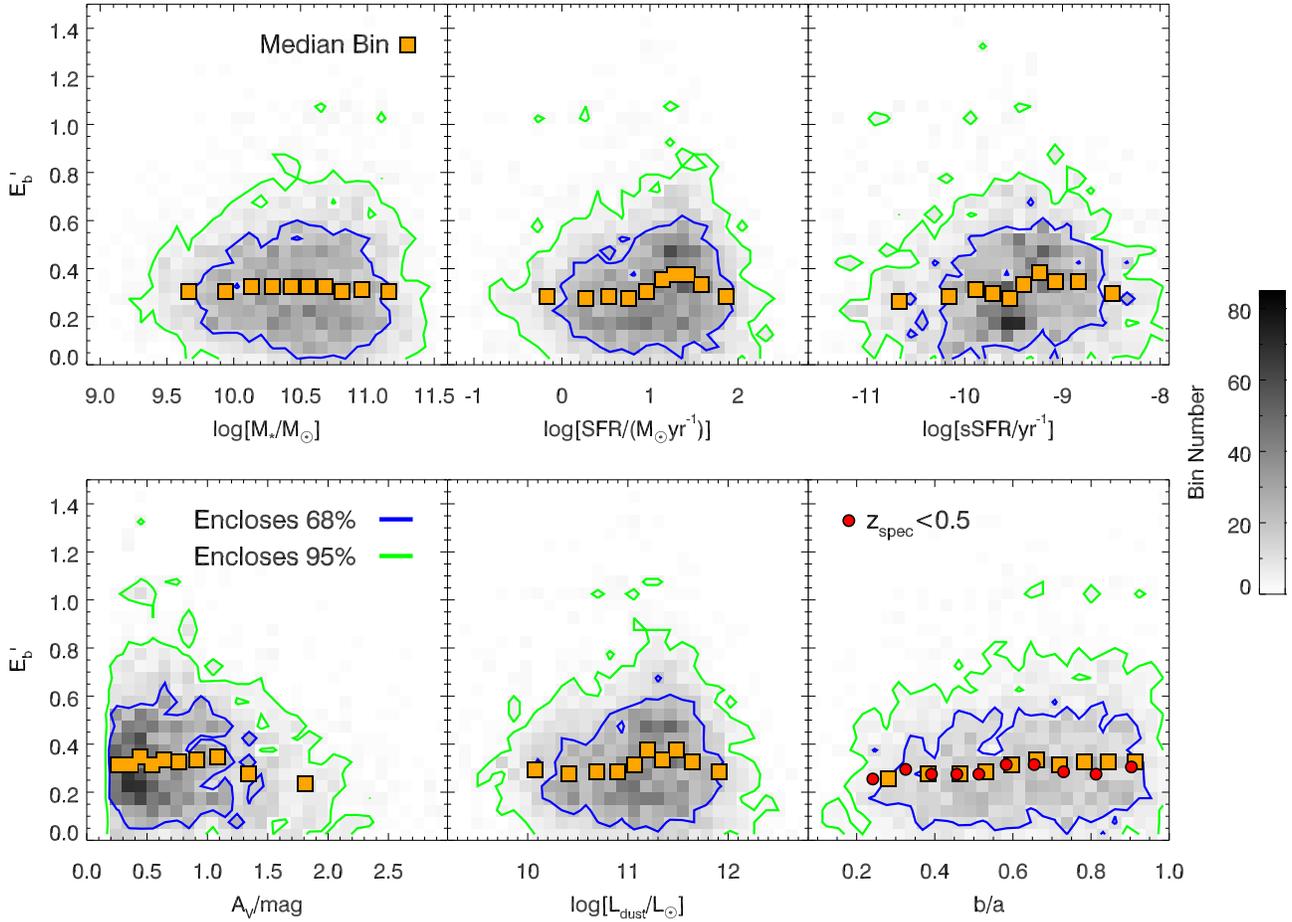} 
\end{center}
\vspace{-0.4cm}
\caption{Similar to Figure~\ref{fig:nCF_2Dhist} but showing the 2175\ang bump strength $E_b'$ as a function of galaxy parameters. No significant trends are observed, however the uncertainties for this quantity are large (see text for more details).
\label{fig:Eb_2Dhist}}
\end{figure*}

Next, we examine the 2175\ang bump strength $E_b'$ as a function of galaxy parameters. The comparisons with galaxy properties are shown in Figure~\ref{fig:Eb_2Dhist}, and do not show any notable trends. However, we attribute the lack of any trends to the high uncertainties on this parameter. This is a result of relying on photometry instead of spectroscopy and also because the total bump attenuation is degenerate with the reddening ($A_{2175}=E_b\cdot E(B-V)$). The confidence ranges for $E_b'$ have values of median$(E_b'(84\%)-E_b'(16\%))=0.30$ and this accounts for a significant fraction of the observed scatter. We also performed this comparison for only galaxies with at least 2 filters in the feature region (same sample used to define the $E_b'$ prior) but did not find significant differences. For now, we conclude only that the median strength for the 2175\ang feature for the entire sample of SFGs is $E_b'=0.31^{+0.20}_{-0.16}$, corresponding to $\sim$30\% of the strength of this feature in the MW extinction curve. As mentioned before, these values are in good agreement with other studies \citep[e.g.,][]{buat12, salim18}.

\begin{figure*}
\begin{center}
\includegraphics[width=0.95\textwidth,clip=true]{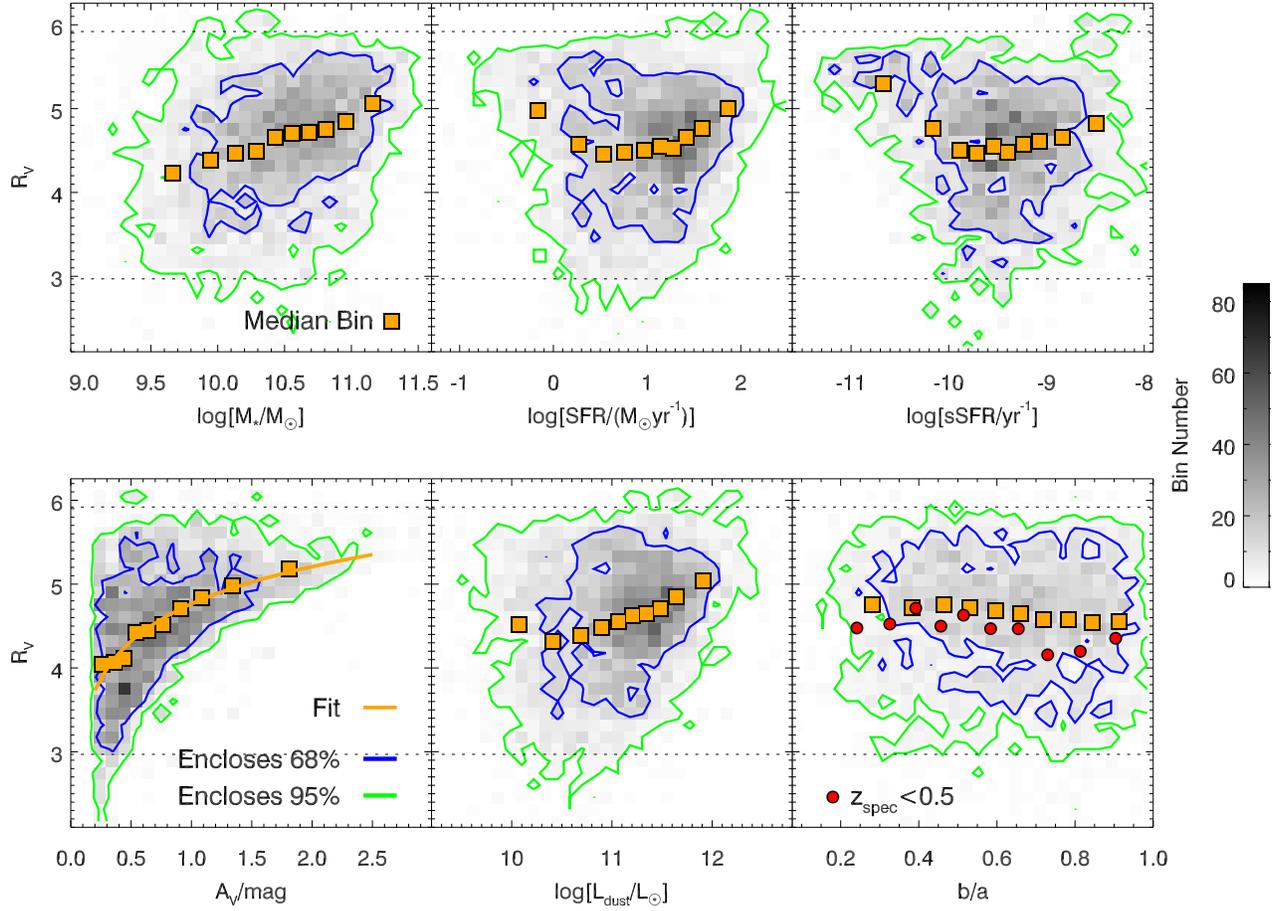}
\end{center}
\vspace{-0.4cm}
\caption{Similar to Figure~\ref{fig:nCF_2Dhist} but showing the total-to-selective attenuation in the $V$-band, $R_V=A_V/E(B-V)$, as a function of galaxy parameters. Marginal trends are evident for $M_*$ and $A_V$. $R_V$ is linked to the value of $n_\mathrm{CF}\mathrm{(optical)}$ and therefore mirrors most trends in Figure~\ref{fig:nCF_2Dhist}. The dotted lines represent the soft boundaries imposed by $n_\mathrm{CF}^{\,\mathrm{ISM}}$ and $n_\mathrm{CF}^{\,\mathrm{BC}}$.
\label{fig:Rv_2Dhist}}
\end{figure*}

Finally, we also examine the total-to-selective attenuation in the $V$-band, $R_V\equiv A_V/E(B-V)$, as a function of galaxy properties in Figure~\ref{fig:Rv_2Dhist}. A commonly adopted assumption when correcting for the effects of dust attenuation is to use the local starburst attenuation curve from \citet{calzetti00} for which $R_V=4.05$. \magphys\ adopts the \citet{charlot&fall00} formalism, where the two dust attenuation components have $\hat{R}_{V}^{\,\mathrm{ISM}}=5.92$ and $\hat{R}_{V}^{\,\mathrm{BC}}=2.97$ for $n_\mathrm{CF}=0.7$ and 1.3, respectively\footnote{These follow from the standard definition of $R_V$, which can be rearranged into $R_V\equiv 1/(\tau_B/\tau_V-1)=1/((\lambda_B/\lambda_V)^{-n_\mathrm{CF}}-1)$ using $A_B/A_V=\tau_B/\tau_V$ and $\tau_\lambda/\tau_V=(\lambda/5500\ang)^{-n_\mathrm{CF}}$. We assume $\lambda_B=4400$\ang and $\lambda_V=5500$\AA .}. Similar to what occurs for the slope of the dust curve, these two values represent only soft boundaries to the range of values for total effective $R_V$. Looking at Figure~\ref{fig:Rv_2Dhist}, slight positive trends are evident for $M_*$ and $A_V$ with $R_V$. A very important caveat to mention regarding the \magphys\ $R_V$ values is that the (fixed) slope of $n_\mathrm{CF}^{\,\mathrm{ISM}}=0.7$ for the diffuse dust component is shallower than most empirically derived attenuation curves at long wavelengths ($\lambda>5500$\AA; see Figure~\ref{fig:att_curve_comparison}, \textit{Left}). We expect that this assumption will impact the derived values of $R_V$ because of the degeneracy between the dust curve slope and normalization for the energy balance. These issues are discussed further in Appendix~\ref{app_nISMvary}. Exploring this in greater detail requires further altering the dust prescription of \magphys\ and is beyond the scope of this paper.

\begin{table*}
\begin{center}
\caption{Summary of dust attenuation curve parameters vs. physical properties relationships and statistics for COSMOS galaxies shown in Figure~\ref{fig:nCF_2Dhist}-\ref{fig:Rv_2Dhist}. These results are for the case of assuming $n_\mathrm{CF}^{\,\mathrm{BC}}=1.3$ and $n_\mathrm{CF}^{\,\mathrm{ISM}}=0.7$ for the two dust components.}
\begin{tabular}{cclcc}
\hline
$y$ & $x$ & fit & Spearman ($\rho$) & Kendall ($\tau$) \\
\hline
$n_\mathrm{CF}\mathrm{(UV)}$ & log[$M_*$] & \nodata & $-0.15$ & $-0.10$  \\
 & log[SFR] & $y=1.09-0.154x$ & $-0.60$ & $-0.42$  \\
 & log[sSFR] & $y=-0.240-0.123x$ & $-0.45$ & $-0.31$  \\
 & $A_V$ & $y=0.883-0.123[\log(x)]-0.327[\log(x)]^2$  & $-0.77$ & $-0.61$  \\
 & log[$L_\mathrm{dust}$] & $y=3.00-0.185x$ & $-0.66$ & $-0.48$  \\
 & $b/a$ & \nodata & $0.06$ & $0.04$  \\
\hline
$E_b'$ & log[$M_*$] & \nodata & $-0.01$ & $-0.01$  \\
 & log[SFR] &  \nodata & $0.08$ & $0.06$  \\
 & log[sSFR] &  \nodata & $0.10$ & $0.06$  \\
 & $A_V$ &  \nodata  & $-0.11$ & $-0.07$  \\
 & log[$L_\mathrm{dust}$] &  \nodata & $0.05$ & $0.03$  \\
 & $b/a$ & \nodata & $0.12$ & $0.08$  \\
\hline
$R_V$ & log[$M_*$] & \nodata & $0.29$ & $0.20$  \\
 & log[SFR] & \nodata & $0.08$ & $0.06$  \\
 & log[sSFR] & \nodata  & $-0.13$ & $-0.08$  \\
 & $A_V$ & $y=4.76+1.48[\log(x)]$  & $0.53$ & $0.39$  \\
 & log[$L_\mathrm{dust}$] & \nodata  & $0.26$ & $0.18$  \\
 & $b/a$ & \nodata & $-0.07$ & $-0.05$  \\
\hline
\hline
\end{tabular}
\label{tab:fit_metrics}
\end{center}
\end{table*}

\section{Conclusion}\label{conclusion}
We extend the spectral modeling code \magphys\ to include a 2175\ang absorption feature and use it to examine the dust attenuation curves of SFGs at $0.1< z\lesssim3$ in the COSMOS field. The main results of the paper are summarized below:
\begin{itemize} 
\item The 2175\ang feature is required in our SED modeling to reduce fitting residuals of this region, with a median strength of $E_b'=0.31^{+0.20}_{-0.16}$ ($\sim$1/3 of the strength in the Milky Way extinction curve).
\item The total effective dust attenuation curve slope shows strong variation with star formation rate and total dust content ($A_V$), such that galaxies with higher SFR and $A_V$ tend to exhibit shallower dust attenuation curves. These are consistent with expectations from radiative transfer theory where attenuation curves become shallower with increasing dust optical depth.
\item We do not find strong variation in the 2175\ang bump strength $E_b'$ with galaxy properties, but this may be driven by the high uncertainty associated with this quantity when using only photometry.
\end{itemize}

We plan to perform a more thorough investigation of dust attenuation curves and the 2175\ang feature using \magphysz\ \citep{battisti19} to examine a much larger sample in the future. We note that the 2175\ang feature accounts for a negligible fraction of the attenuated energy and we find that its inclusion has little impact on the derived physical property estimates (e.g., $M_*$, SFR) when the redshift of the galaxy is known. However, accounting for the feature is very important when attempting to constrain the photo-$z$ of very dusty galaxies using their SEDs \citep{battisti19}, which, in turn, does directly affect the derived values of physical properties that are luminosity-dependent. 

\section*{Acknowledgments} The authors thank the anonymous referee, whose suggestions helped to clarify and improve the content of this work. EdC gratefully acknowledges the Australian Research Council as the recipient of a Future Fellowship (project FT150100079). Parts of this research were supported by the Australian Research Council Centre of Excellence for All Sky Astrophysics in 3 Dimensions (ASTRO 3D), through project number CE170100013. AJB thanks K. Grasha for comments that improved the paper and C. Laigle for correspondence regarding COSMOS2015 catalog usage. AJB is also thankful for attending ASTRO 3D writing retreats that provided a helpful environment to complete portions of this manuscript. We acknowledge the invaluable labor of the maintenance and clerical staff at our institutions, whose contributions make our scientific discoveries a reality. Data analysis made use of Topcat \citep{taylor05}. Based on data products from observations made with ESO Telescopes at the La Silla Paranal Observatory under ESO programme ID 179.A-2005 and on data products produced by TERAPIX and the Cambridge Astronomy Survey Unit on behalf of the UltraVISTA consortium.


\bibliography{AJB_bib}
\clearpage

\appendix
\section{Effect of Allowing \texorpdfstring{$\lowercase{n}_\mathrm{ISM}$}{nISM} to Vary}\label{app_nISMvary}
An important caveat in the dust attenuation curves that \magphys\ infers is that they are dependent on the assumptions of the \citet{charlot&fall00} dust attenuation prescriptions. In particular, the value of $n_\mathrm{CF}^{\,\mathrm{ISM}}=0.7$ acts as a soft boundary for the minimum shallowness for which the total effective dust curve can have. Several other studies suggest that highly obscured galaxies (large $A_V$) show a preference for dust curve slopes that are even shallower, with $n_\mathrm{CF}<0.7$ \citep[e.g.,][]{chevallard13, salmon16, loFaro17, buat18}. Here we explore the effect on the inferred attenuation curves if $n_\mathrm{CF}^{\,\mathrm{ISM}}$ is free to vary from 0.4 to 1.0 with a flat prior over this range. All other assumptions/priors remain unchanged with respect to the version of \magphys\ high-$z$ described in Section~\ref{magphys_method}.

Similar to the main analysis, we restrict our sample to galaxies for which $A_V>0.2$ and $\chi_\mathrm{red}^2<\bar{\chi}_\mathrm{red}^2+4\sigma(\chi_\mathrm{red}^2)$, where $\bar{\chi}_\mathrm{red}^2$ and $\sigma(\chi_\mathrm{red}^2)$ are the mean and dispersion of a Gaussian fit to the lower 90\% of the $\chi_\mathrm{red}^2$ population (determined separately for G10 and C15+SD), that are considered to have more reliably derived attenuation curves. Using this version, the fits of the $\chi_\mathrm{red}^2$ distribution give $\bar{\chi}_\mathrm{red,G10}^2=0.88$ and $\sigma(\chi_\mathrm{red,G10}^2)=0.40$, and $\bar{\chi}_\mathrm{red,C15+SD}^2=0.50$ and $\sigma(\chi_\mathrm{red,C15+SD}^2)=0.19$. The $A_V$ and $\chi_\mathrm{red}^2$ cuts remove 551 and 505 galaxies (43 in both), respectively, leaving 4770 galaxies for the analysis.

\begin{figure*}
\begin{center}
\begin{minipage}{0.6\textwidth}
\includegraphics[width=\textwidth]{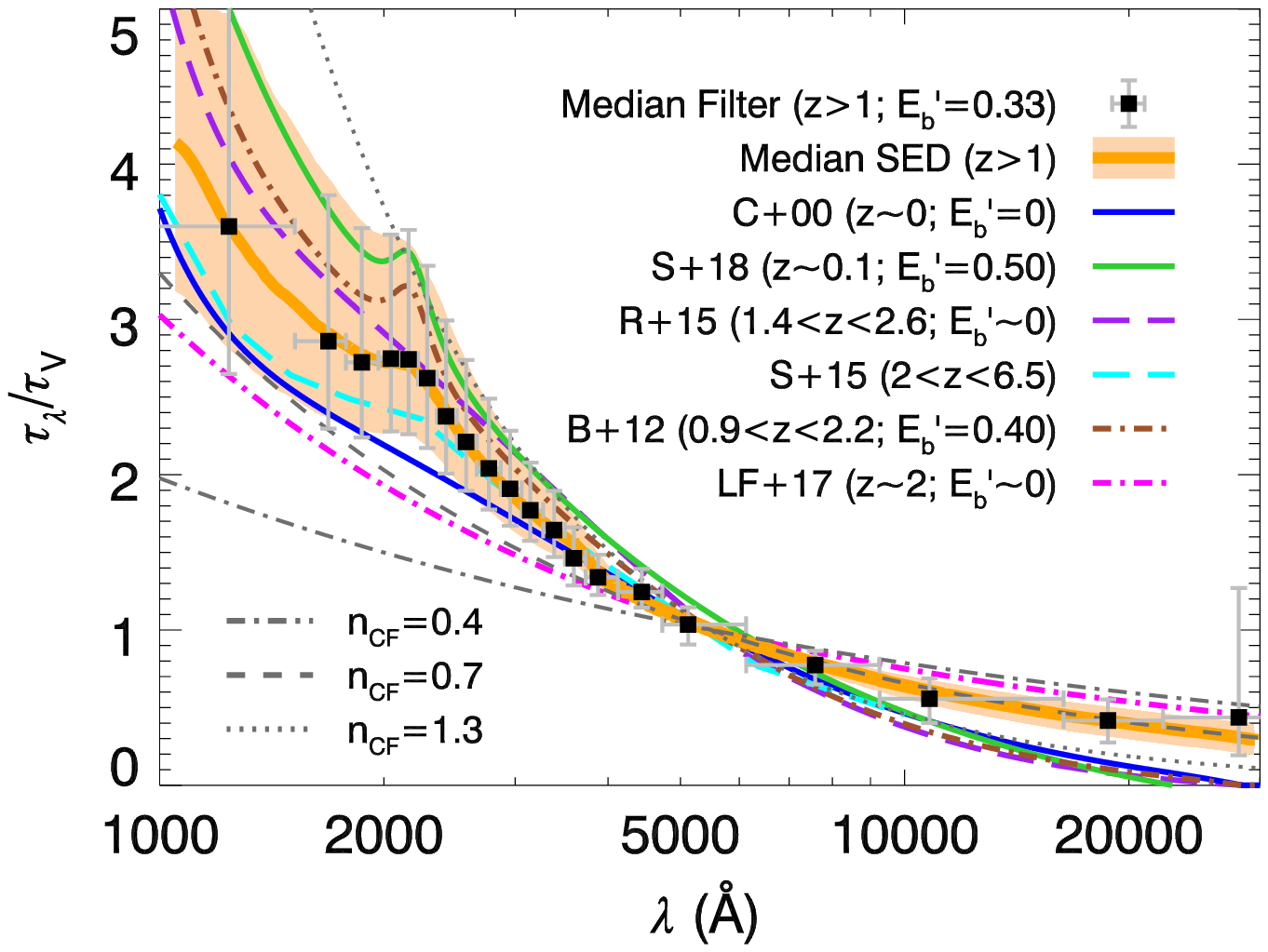}
\end{minipage}
\hspace{1mm}
\begin{minipage}{0.35\textwidth}
\includegraphics[width=\textwidth,clip=true]{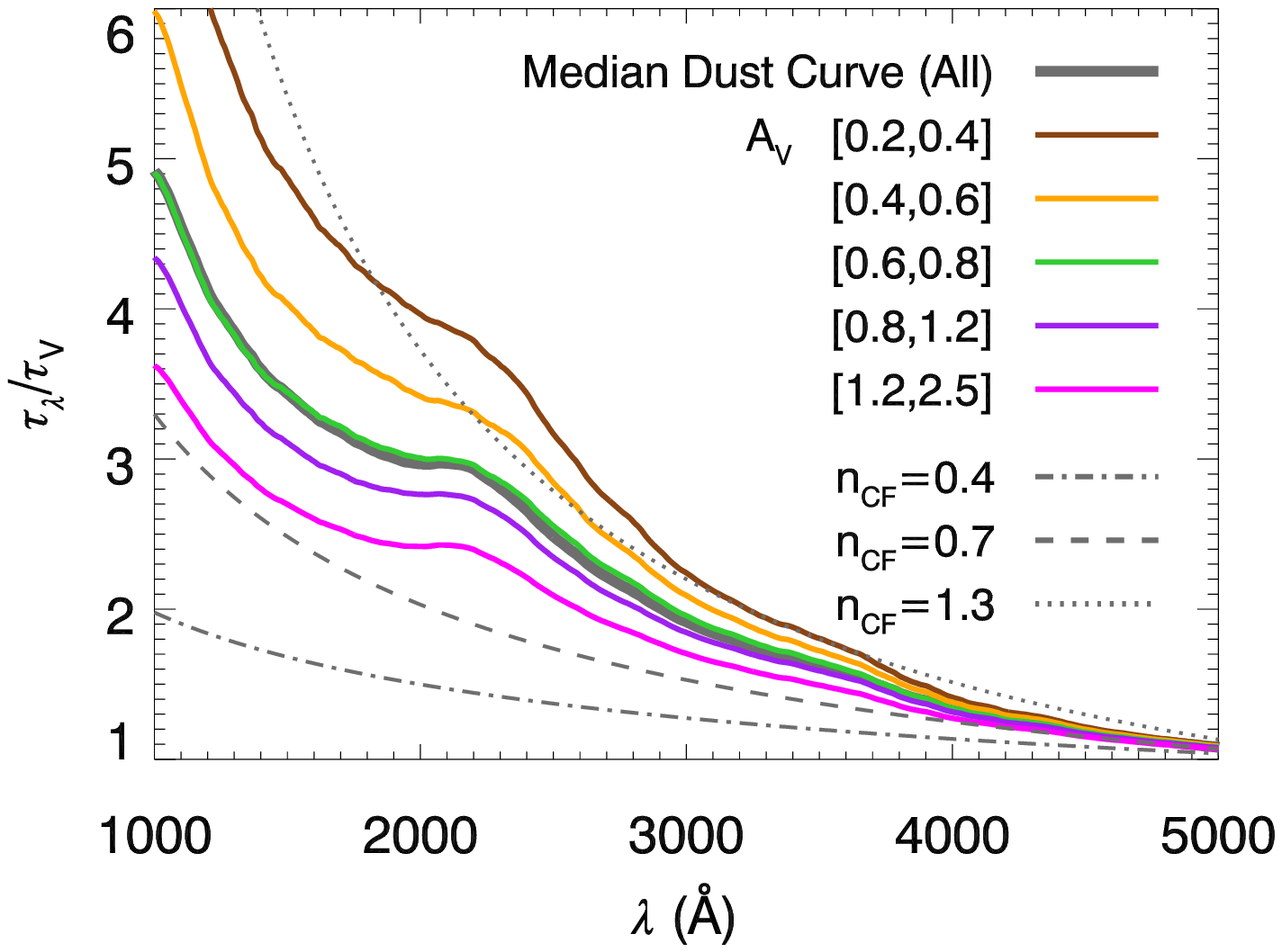} \\
\includegraphics[width=\textwidth]{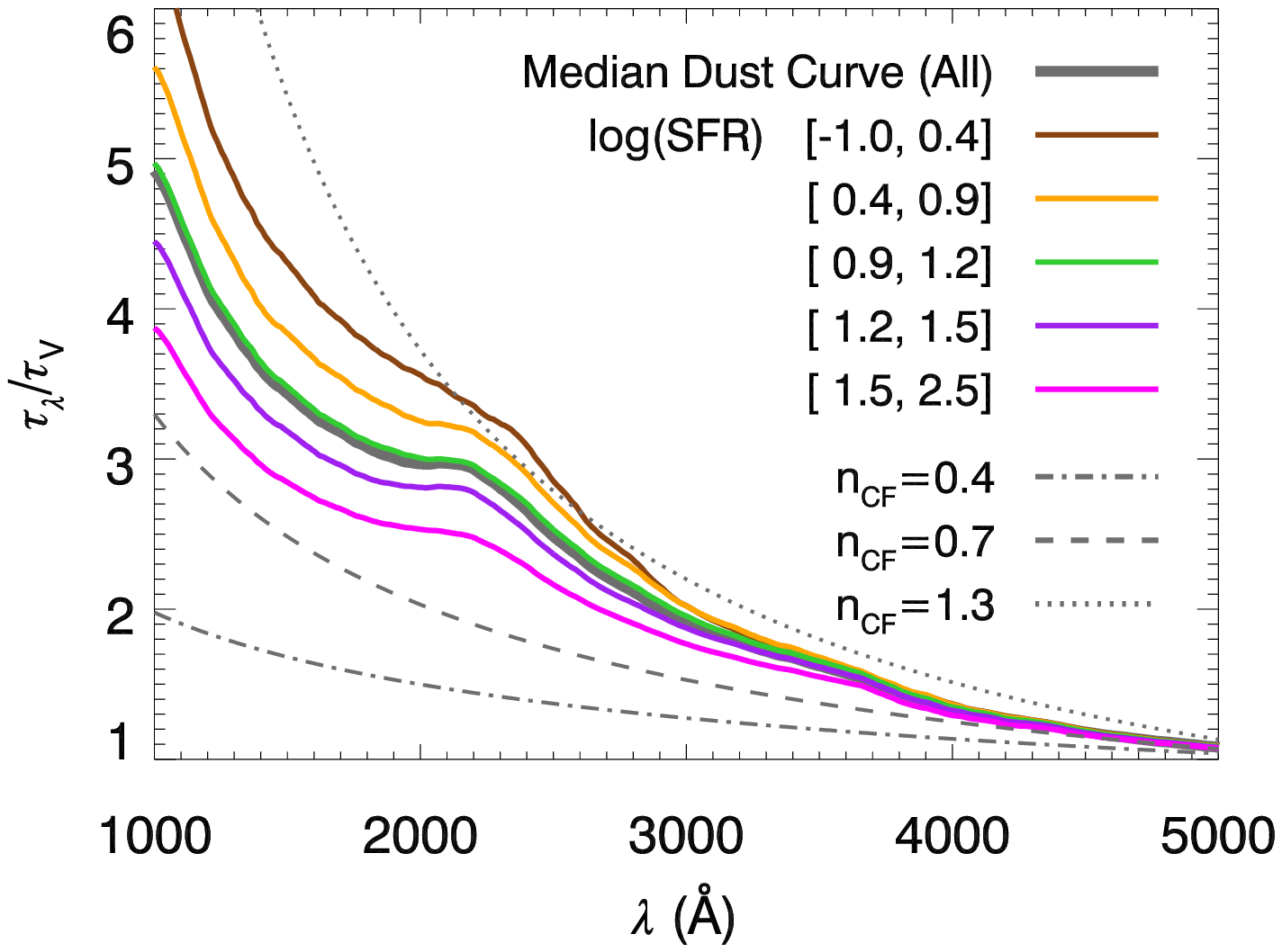}
\end{minipage}
\end{center}
\vspace{-0.3cm}
\caption{Similar to Figure~\ref{fig:dust_curves} but for the case when $n_\mathrm{CF}^{\,\mathrm{ISM}}$ is free to vary from 0.4 to 1.0 (adopt flat prior). The average curve (\textit{left}) is similar to before, but slightly shallower and with a higher dispersion. The main difference from the previous results, assuming a fixed $n_\mathrm{CF}^{\,\mathrm{ISM}}=0.7$, is that now the highest $A_V$ (or SFR) galaxies have shallower inferred attenuation curves (\textit{right panels}).
\label{fig:dust_curves_nISMvary}}
\end{figure*}

The average attenuation curve that is inferred for the $z\geq1$ sample is shown in Figure~\ref{fig:dust_curves_nISMvary}, \textit{Left}, with other curves in the literature also shown for reference. Relative to the average curve shown in Figure~\ref{fig:dust_curves}, this curve is slightly shallower and has a larger variance (orange shaded region) due to the additional flexibility from letting $n_\mathrm{CF}^{\,\mathrm{ISM}}$ vary. The main differences with respect to before occur for galaxies with large $A_V$ and SFR, which show a preference toward shallower total effective dust attenuation curves (Figure~\ref{fig:dust_curves_nISMvary}, \textit{Right}). There is no change on the distribution of the 2175\ang feature, with $E_b'=0.32^{+0.19}_{-0.16}$, and we attribute this lack of change to the fact that this is a normalized bump strength (i.e., $E_b' = E_b/R_V^{\,\mathrm{ISM}}$).

\begin{figure*}
\begin{center}
$\begin{array}{c}
\includegraphics[width=0.95\textwidth,clip=true]{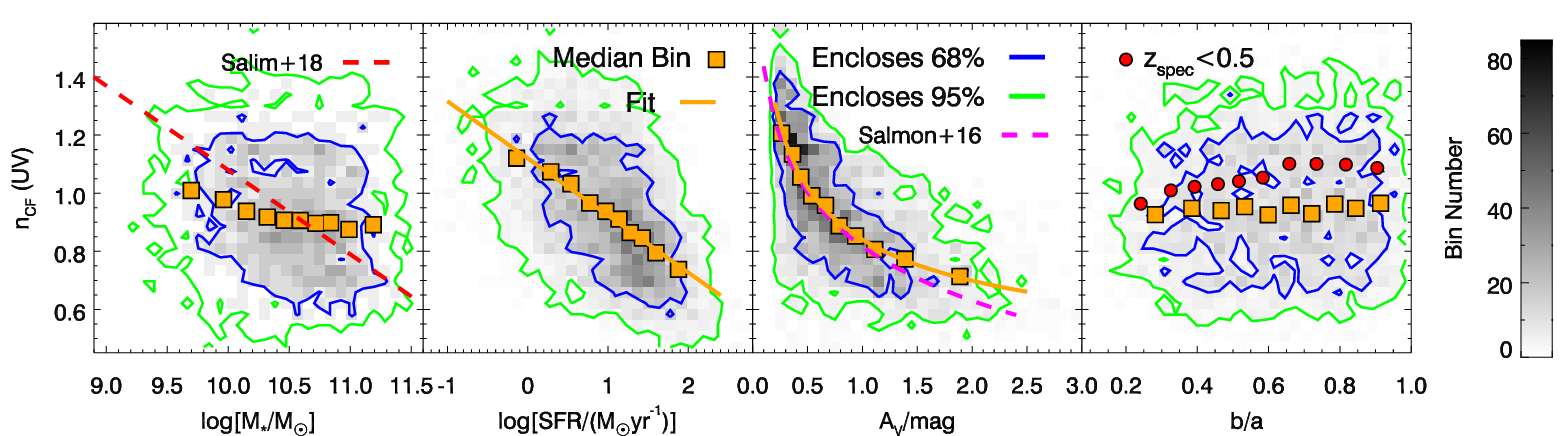}  \\
\includegraphics[width=0.95\textwidth,clip=true]{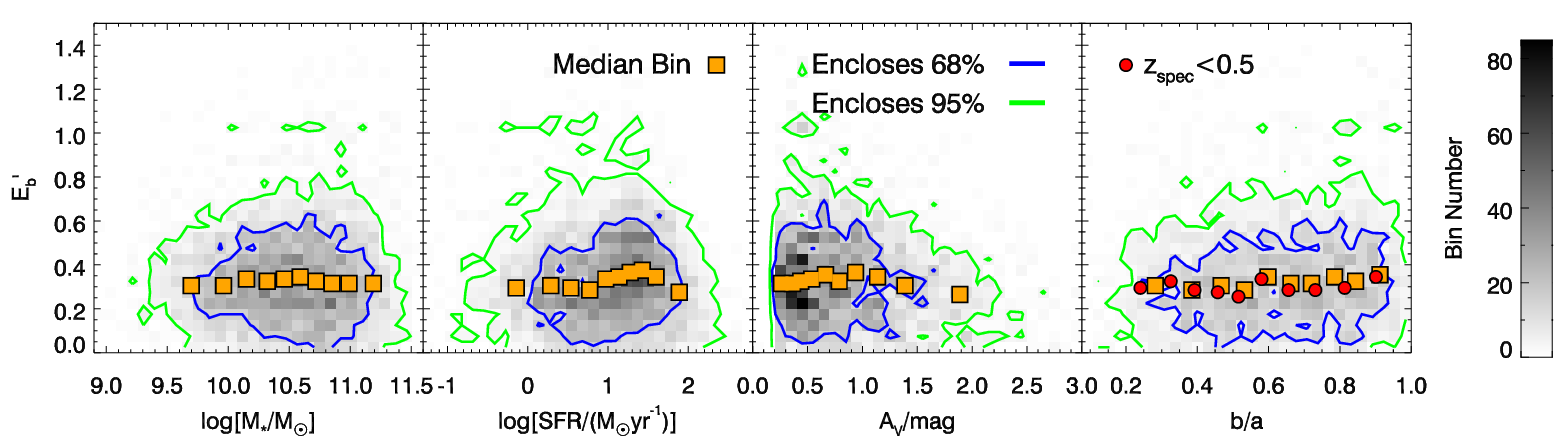} \\
\includegraphics[width=0.95\textwidth,clip=true]{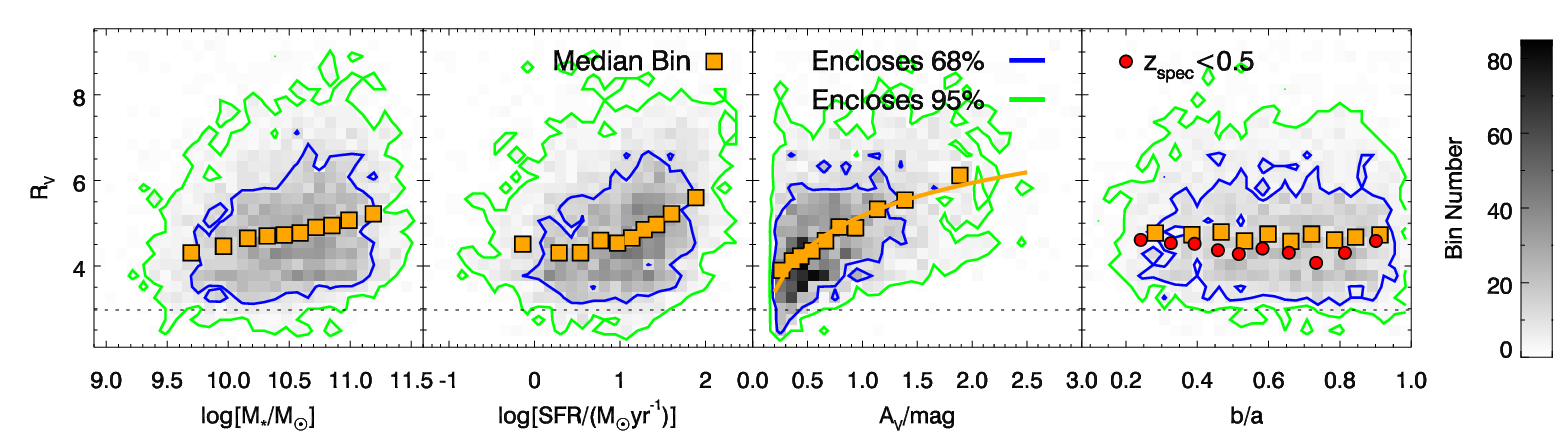} \\
\end{array}$
\end{center}
\vspace{-0.4cm}
\caption{Similar to Figures~\ref{fig:nCF_2Dhist}-\ref{fig:Rv_2Dhist} but now for the case when $n_\mathrm{CF}^{\,\mathrm{ISM}}$ is free to vary from 0.4 to 1.0 (adopt flat prior). Under this assumption, the allowed range of $n_\mathrm{CF}\mathrm{(UV)}$ extends to lower values (shallower curves) and this results in closer agreement to the \citet{salmon16} relation. The allowed range of $R_V$  for this case is also larger because it is linked to the value of $n_\mathrm{CF}\mathrm{(optical)}$. Panels for sSFR and $L_\mathrm{dust}$ are not shown, but behave similarly to the SFR. The dotted line in the lower panel represents the soft boundary imposed by $n_\mathrm{CF}^{\,\mathrm{BC}}=1.3$.
\label{fig:nCF_Eb_2Dhist_nISMvary}}
\end{figure*}

In Figure~\ref{fig:nCF_Eb_2Dhist_nISMvary}, we show the distribution of $n_\mathrm{CF}\mathrm{(UV)}$, $E_b'$, and $R_V$ as a function of various galaxy properties. A summary of the Spearman $\rho$ and Kendall $\tau$ rank correlation coefficients and functional fits, when appropriate, for these parameters are shown in Table~\ref{tab:fit_metrics_nISMvary}. The main differences relative to Figure~\ref{fig:nCF_2Dhist}-\ref{fig:Rv_2Dhist} are that the relations with $A_V$ and SFR (also sSFR and $L_\mathrm{dust}$; not shown) are steeper as a result of the allowed curve slope range being larger. The $n_\mathrm{CF}\mathrm{(UV)}$-$A_V$ relation is more consistent with the relation from \citet{salmon16} than when using a fixed $n_\mathrm{CF}^{\,\mathrm{ISM}}=0.7$.

\begin{figure}
\begin{center}
\includegraphics[width=0.46\textwidth]{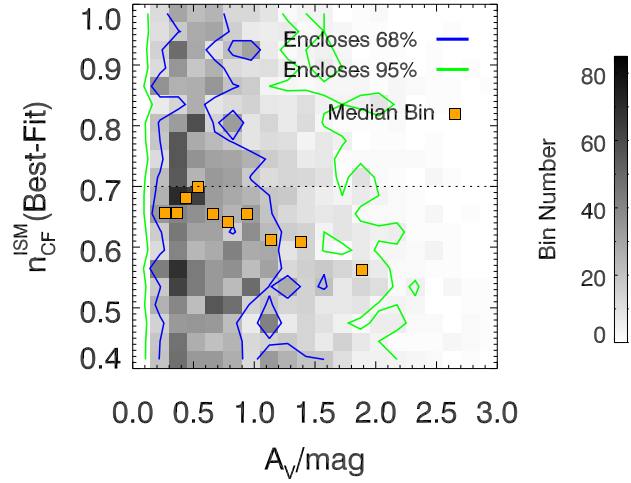}
\end{center}
\vspace{-0.3cm}
\caption{Comparison between the $n_\mathrm{CF}^{\,\mathrm{ISM}}$ values for the best-fit SEDs as a function of $A_V$. $A_V$ shows the strongest trend with the slope of the inferred total effective dust attenuation curve ($n_\mathrm{CF}\mathrm{(UV)}$) but there doesn't appear to be a clear preference on the choice of $n_\mathrm{CF}^{\,\mathrm{ISM}}$, except perhaps for $A_V\gtrsim1$~mag where median begins to decrease. The poor correlation here highlights that there is significant degeneracy in the manner that the various model parameters can be combined to reproduce the observed galaxy SED.
\label{fig:nISM_Av_comparison}}
\end{figure}

Interestingly, if we look at the distribution of $n_\mathrm{CF}^{\,\mathrm{ISM}}$ for the best-fit SEDs as a function of $A_V$, shown in Figure~\ref{fig:nISM_Av_comparison}, there appears to be almost no preference on the value of $n_\mathrm{CF}^{\,\mathrm{ISM}}\textrm{(Best-fit)}$, except perhaps for $A_V\gtrsim1$ where the median begins to decrease with increasing $A_V$. We attribute this lack of correlation to the significant degeneracy in the manner that the various model parameters can be combined to reproduce the observed galaxy SED. The median value of $n_\mathrm{CF}^{\,\mathrm{ISM}}\textrm{(Best-fit)}$ is 0.64 with a dispersion of 0.16.

We conclude that the assumptions on the shape of the birth-cloud and ISM dust components in \magphys\ only significantly impact the slope of the inferred total effective dust attenuation curve for highly obscured galaxies, where there appears to be a preference for shallower curves relative to the standard $n_\mathrm{CF}^{\,\mathrm{ISM}}=0.7$ in the \citet{charlot&fall00} formalism \citep{chevallard13, salmon16, loFaro17, buat18}. However, we note that there is significant degeneracy in how to account for this when utilizing two dust components and that the methods adopted in this section are not unique.

     
\clearpage

\begin{table*}
\begin{center}
\caption{Summary of dust attenuation curve parameters vs. physical properties relationships and statistics for COSMOS galaxies shown in Figure~\ref{fig:nCF_Eb_2Dhist_nISMvary}. These results are for the case of assuming $n_\mathrm{CF}^{\,\mathrm{BC}}=1.3$ and $0.4\leq n_\mathrm{CF}^{\,\mathrm{ISM}}\leq 1.0$ (flat prior) for the two dust components.}
\begin{tabular}{cclcc}
\hline
$y$ & $x$ & fit & Spearman ($\rho$) & Kendall ($\tau$) \\
\hline
$n_\mathrm{CF}\mathrm{(UV)}$ & log[$M_*$] & \nodata & $-0.15$ & $-0.10$  \\
 & log[SFR] & $y=1.12-0.197x$ & $-0.54$ & $-0.37$  \\
 & log[sSFR] & $y=-0.514-0.151x$ & $-0.39$ & $-0.27$  \\
 & $A_V$ & $y=0.841-0.533[\log(x)]-0.203[\log(x)]^2$  & $-0.67$ & $-0.50$  \\
 & log[$L_\mathrm{dust}$] & $y=3.47-0.229x$ & $-0.58$ & $-0.40$  \\
 & $b/a$ & \nodata & $0.05$ & $0.03$  \\
\hline
$E_b'$ & log[$M_*$] & \nodata & $0.01$ & $0.01$  \\
 & log[SFR] &  \nodata & $0.05$ & $0.03$  \\
 & log[sSFR] &  \nodata & $0.06$ & $0.04$  \\
 & $A_V$ &  \nodata  & $-0.09$ & $-0.06$  \\
 & log[$L_\mathrm{dust}$] &  \nodata & $0.03$ & $0.02$  \\
 & $b/a$ & \nodata & $0.11$ & $0.07$  \\
\hline
$R_V$ & log[$M_*$] & \nodata & $0.21$ & $0.14$  \\
 & log[SFR] & \nodata & $0.26$ & $0.17$  \\
 & log[sSFR] & \nodata  & $0.09$ & $0.06$  \\
 & $A_V$ & $y=5.17+2.55[\log(x)]$  & $0.49$ & $0.35$  \\
 & log[$L_\mathrm{dust}$] & \nodata  & $0.33$ & $0.23$  \\
 & $b/a$ & \nodata & $-0.04$ & $-0.03$  \\
\hline
\hline
\end{tabular}
\label{tab:fit_metrics_nISMvary}
\end{center}
\end{table*}

\end{document}